\documentclass[12pt]{iopart}
\usepackage{iopams}  
\usepackage{graphicx}
\usepackage{color}

\newcommand{\rin}{r_\mathrm{in}}
\newcommand{\aef}{\alpha_\mathrm{EF}}
\newcommand{\sos}{c_\mathrm{s}}
\newcommand{\sosi}{c_\mathrm{s, \infty}}
\newcommand{\soss}{c_\mathrm{s, \ast}}

\begin{document}

\title[Relativistic Low Angular Momentum Accretion]{Relativistic Low Angular Momentum Accretion: Long Time Evolution of Hydrodynamical Inviscid Flows}

\author{Patryk Mach}
\address{Institute of Physics, Jagiellonian University, \L{}ojasiewicza 11, 30-438 Krak\'{o}w, Poland}
\ead{patryk.mach@uj.edu.pl}

\author{Micha{\l} Pir\'{o}g}
\address{Institute of Physics, Jagiellonian University, \L{}ojasiewicza 11, 30-438 Krak\'{o}w, Poland}
\ead{michal.pirog@uj.edu.pl}

\author{Jos\'{e} A. Font}
\address{Departamento de Astronomia y Astrof\'{\i}sica, Universitat de Val\`encia, Dr.~Moliner, 50, 46100 - Burjassot (Val\`encia), Spain}
\address{Observatori Astron\`omic, Universitat de Val\`encia, C/ Catedr\'atico 
  Jos\'e Beltr\'an 2, 46980, Paterna (Val\`encia), Spain} 

\begin{abstract}
We investigate relativistic low angular momentum accretion of inviscid perfect fluid onto a Schwarzschild black hole. The simulations are performed with a general-relativistic, high-resolution (second-order), shock-capturing, hydrodynamical numerical code. We use horizon-penetrating Eddington-Finkelstein coordinates to remove inaccuracies in regions of strong gravity near the black hole horizon and show the expected convergence of the code with the Michel solution and stationary Fishbone-Moncrief toroids. We recover, in the framework of relativistic hydrodynamics, the qualitative behavior known from previous Newtonian studies that used a Bondi background flow in a pseudo-relativistic gravitational potential with a latitude-dependent angular momentum at the outer boundary. Our models exhibit characteristic ``turbulent'' behavior and the attained accretion rates are lower than those of the Bondi--Michel radial flow. For sufficiently low values of the asymptotic sound speed, geometrically thick tori form in the equatorial plane surrounding the black hole horizon while accretion takes place mainly through the poles. 
\end{abstract}

\pacs{04.20.-9, 98.62.Mw}

\maketitle

\section{Introduction}
\label{sec_intro}

Low angular momentum accretion models have been proposed in order to explain a potential paradox connected with the existence of inactive galactic centers (see \cite{proga_begelman} and references therein). On the one hand it is believed that many galaxies contain supermassive black holes in their centers, and there are good reasons to assume that there is a lot of matter available for accretion in the central regions of galaxies. On the other hand only a few of the observed galaxies can be classified as ``active''. Thus, it seems important to search for accretion models characterized by low accretion rates. Low angular momentum accretion models belong to this class.

There are many analytic approximate models of low angular momentum hydrodynamical accretion (both Newtonian and relativistic) constructed with the aim to obtain stationary solutions (examples and references can be found e.g.\ in Ref.\ \cite{das}; see also \cite{c, b}). There is however (to the best of our knowledge) no exact analytic, general model (solution) even for a much simpler case of radially dominated Bondi-type accretion on the Kerr spacetime (with the exception of a simple model for the ultrahard fluid \cite{petrich_shapiro} or approximate models similar to the one introduced in Ref.\ \cite{tejeda}).

In this work we perform general relativistic numerical simulations of low angular momentum accretion onto Schwarzschild black holes, following the approach introduced by Proga and Begelman \cite{proga_begelman} who first investigated numerically this type of accretion flows employing Newtonian simulations (but see also \cite{abramowicz_zurek, a}). In order to mimic relativistic effects the simulations of \cite{proga_begelman} made use of the Paczy\'{n}ski--Wiita (pseudo-Newtonian) gravitational potential \cite{paczynski_wiita} which for a Schwarzschild black hole captures general relativity at the 10--20\% level outside the marginally stable orbit. In our work we neglect the self-gravity of the fluid, which is assumed to evolve in the fixed curved background provided by the Schwarzschild metric. The numerical model presented in Ref.\ \cite{proga_begelman} is based upon the following simple idea: Consider, as a background solution, spherically symmetric stationary radial Bondi flow of a perfect gas onto a central, massive black hole \cite{bondi}. Such a solution can be easily obtained numerically. Then add some angular momentum, so that the azimuthal component of the velocity $v_\phi$ is no longer zero. This can be done almost arbitrarily, and thus the choices of the dependence of the angular momentum on the coordinates may seem, at first, to be completely ad hoc. The initial data prepared in this way are then evolved numerically in time.  An important factor that affects the evolution of the initial models is the way in which the boundary conditions are implemented. The conditions at the inner boundary of the grid are almost straightforward. In the Newtonian case one simply assumes outflowing boundary conditions, appropriate for matter flowing out of the grid towards the black hole. In a general-relativistic setup, given the right choice of coordinates, one can make sure that the inner radial boundary of the grid is located within the black hole horizon so that its influence on the fluid evolution is causally deactivated -- that actually constitutes one of the major advantages of working in the relativistic framework. The subsequent hydrodynamical evolution will thus only depend on the conditions that are implemented at the outer boundary of the grid, and it is important to make this dependence as minimal as possible. The natural way to achieve this is to place the outer boundary as far from the central black hole as possible. This comes at the price of increasing the computational cost. In Ref.\ \cite{proga_begelman} the outer boundary of the numerical grid is placed outside the Bondi radius -- the characteristic length scale associated with the Bondi solution. The boundary conditions are implemented so that the angular momentum is effectively injected into the numerical grid, but the mass accretion rate can vary with time. 

The first axisymmetric and purely hydrodynamical simulations of Proga and Begelman~\cite{proga_begelman} showed that the material with an excess of angular momentum  forms a geometrically thick torus near the equator, which constrains and significantly reduces the accretion rates through the polar regions. These simulations were later extended to include magnetic fields \cite{proga_begelman_magneto} and three-dimensional computations \cite{janiuk_proga_kurosawa}. In Ref.\ \cite{janiuk_sznajder_moscibrodzka_proga} the dependence of the results on the adiabatic index and on the  boundary conditions were also discussed.

There are numerous similar relativistic simulations in the test-fluid regime. Most of the existing ones that deal with geometrically thick accretion configurations assume an already formed thick disk as initial data, e.g.~\cite{font_daigne1, font_daigne2}. The exceptions include Refs.\ \cite{bambi1, bambi2, bambi3}, in which accretion flows onto Kerr black holes are investigated. In the majority of these simulations the outer boundary of the numerical grid is placed relatively close to the (outer) horizon of the Kerr black hole. Here, in order to generalize the Newtonian simulations of Ref.\ \cite{proga_begelman} to the relativistic case, the outer boundary is placed at 1.5 of the Bondi radius. We note that in Ref.\ \cite{Fragile}, the authors discuss a model that is very similar to the one investigated in this paper as a test for their numerical code.

As we show below, the results of our relativistic simulations confirm previous Newtonian results at the qualitative level. The evolution of the various models considered depends strongly on the asymptotic value of the speed of sound $\sosi$; for those with small enough $\sosi$, a geometrically thick torus forms around the central black hole. These tori also exhibit a ``turbulent'' hydrodynamical behavior. In all cases the observed accretion rate drops to about one third of the value characteristic for the unperturbed, spherically symmetric Michel flow (which is already quite low).

We note that the terms ``turbulent'' and ``turbulence'' are used in this paper in square quotes. ``Turbulence'' is not understood here in any strict mathematical sense. We use the expression ``turbulent behavior'' in reference to a chaotic motion that follows the symmetry breaking of the simulated flow.

The organization of this paper is as follows. In Sec.\ \ref{sec_code} we describe our numerical code, that was adapted from the one used in Ref.\ \cite{font_bondi_hoyle}. Section \ref{sec_tests} describes exact solutions that we use to test new aspects of the numerical code and that are specific to the accretion problem: the Michel solution expressed in Eddington--Finkelstein coordinates and stationary toroids obtained by Fishbone and Moncrief (also in Eddington--Finkelstein coordinates). In Sec.\ \ref{sec_ic} we discuss the initial and boundary conditions that are used in the presented simulations. An overview of the results of our numerical study can be found in Sec.\ \ref{sec_results}. Finally, the last section contains a summary of our work.

\section{Numerical code}
\label{sec_code}

The numerical code used for the simulations discussed in this paper is adapted from the one used in Ref.\ \cite{font_bondi_hoyle} to simulate relativistic Bondi--Hoyle accretion. The equations of hydrodynamics are derived in horizon-penetrating Eddington--Finkelstein spherical coordinates $(t,r,\theta,\phi)$. In these coordinates the Schwarzschild metric has the form (in $c = G = 1$ units; such geometrized units are used throughout this paper)
\begin{eqnarray}
\nonumber
ds^2 & = &  - \left( 1 - \frac{2M}{r} \right) dt^2 + \frac{4M}{r} dt dr + \left( 1 + \frac{2M}{r} \right) dr^2 \\
& & + r^2 \left( d \theta^2 + \sin^2 \theta d\phi^2 \right).
\label{zzzc}
\end{eqnarray}
The line element (\ref{zzzc}) can be written in the standard 3+1 splitting form, that is
\begin{eqnarray}
ds^2 = - (\alpha^2 - \beta_i \beta^i) dt^2 + 2 \beta_i dx^i dt + \gamma_{ij} dx^i dx^j,  
\end{eqnarray}
with $\beta_r = 2M/r$, $\alpha = \sqrt{r/(r + 2M)}$, $\gamma_{rr} = 1 + 2M/r$, $\gamma_{\theta\theta} = r^2$, $\gamma_{\phi \phi} = r^2 \sin^2 \theta$.

The equations of hydrodynamics for a perfect fluid stress-energy tensor read:
\begin{eqnarray}
\label{zzza}
\nabla_\mu (\rho u^\mu) = 0\,,
\end{eqnarray}
and
\begin{eqnarray}
\label{zzzb}
\nabla_\mu T^{\mu \nu} \equiv \nabla_\mu \left( \rho h u^\mu u^\nu +  p g^{\mu \nu} \right) = 0\,.
\end{eqnarray}
Here $\rho$ is the rest-mass density, $h = 1 + \epsilon + p/\rho$ is the specific enthalpy, $p$ is the pressure, and $u^\mu$ denotes the four velocity of the fluid. The specific internal energy is denoted with $\epsilon$. We use the perfect gas equation of state $p = (\Gamma - 1) \rho \epsilon$, where $\Gamma$ is a constant, although generalization to the form $p = p(\rho,\epsilon)$ is straightforward.

Equations (\ref{zzza}) and (\ref{zzzb}) are written as an explicitly hyperbolic set of conservation laws following the so-called ``Valencia'' formulation \cite{valencia}. Three-velocities of the fluid are defined as
\begin{eqnarray}
v^i = \frac{u^i}{\alpha u^0} + \frac{\beta^i}{\alpha}, \hspace{0.5cm} (i=r,\theta,\phi)
\end{eqnarray}
and the Lorenz factor is $W = \alpha u^0$. With these conventions the following standard relation holds:
\begin{eqnarray}
W^2 = 1/\left( 1 - v_i v^i \right). 
\end{eqnarray}

The equations of hydrodynamics can be written in flux-conservative form
\begin{eqnarray}
\frac{1}{\sqrt{-g}} \partial_t \left( \sqrt{\gamma} \mathbf q \right) + \frac{1}{\sqrt{-g}} \partial_i \left( \sqrt{-g} \mathbf F^i \right) = \mathbf \Sigma\, ,
\end{eqnarray} 
where $g = \det g_{\mu \nu}$, $\gamma = \det \gamma_{ij}$. Note that $\sqrt{-g} = \alpha \sqrt{\gamma}$. Here $\mathbf q$ is a vector of conserved quantities
\begin{eqnarray}
\mathbf q = \left[ D, S_j, \tau \right] \equiv \left[ \rho W, \rho h W^2 v_j, \rho h W^2 - p - \rho W \right]\,,
\end{eqnarray} 
$\mathbf F^i$ denote flux vectors
\begin{eqnarray}
\mathbf F^i & = & \left[ D \left(v^i - \frac{\beta^i}{\alpha} \right), S_j \left( v^i - \frac{\beta^i}{\alpha} \right) + p \delta_j^i, \right. \left. \tau \left(v^i - \frac{\beta^i}{\alpha} \right) + p v^i \right]\,,
\end{eqnarray}
and $\mathbf \Sigma$ is a vector of source terms
\begin{eqnarray}
\mathbf \Sigma & = & \left[0, T^{\mu \nu} \left( \partial_\mu g_{\nu j} - \Gamma^\delta_{\mu \nu} g_{\delta j} \right), \right.  \left. \alpha \left( T^{\mu 0} \partial_\mu \ln \alpha - T^{\mu \nu} \Gamma^0_{\mu \nu} \right) \right]\,.
\end{eqnarray}
It is convenient to write the above system of equations as
\begin{eqnarray}\partial_t \mathbf q + \partial_i \left( \alpha \mathbf F^i \right) = \alpha \mathbf \Sigma - \frac{\partial_i \sqrt{\gamma}}{\sqrt{\gamma}} \alpha \mathbf F^i\,.
\end{eqnarray}
In Eddington--Finkelstein coordinates, this yields
\begin{eqnarray}
\partial_t \mathbf q & + & \partial_r \left( \alpha \mathbf F^r \right) + \partial_\theta \left( \alpha \mathbf F^\theta \right) + \partial_\phi \left( \alpha \mathbf F^\phi \right) = \alpha \mathbf \Sigma \nonumber \\
& - & \left( 2 - \frac{M \alpha^2}{r} \right) \frac{\alpha}{r} \mathbf F^r - \cot \theta \alpha \mathbf F^\theta\,.
\end{eqnarray}
A simple but lengthy calculation allows one to establish the components of $\mathbf \Sigma$ in Eddington--Finkelstein coordinates. One obtains
\begin{eqnarray}
\Sigma_1 & = & 0, \\
\Sigma_2 & = & \frac{2p}{r} + \frac{\rho h W^2}{r} \left[ v_\theta v^\theta + v_\phi v^\phi - \frac{M}{r} (v^r + \alpha)^2 \right], \\
\Sigma_3 & = & \cot \theta \left( p + \rho h W^2 v_\phi v^\phi \right), \\
\Sigma_4 & = & 0, \\
\Sigma_5 & = & \frac{M \alpha}{r^2} \left\{ 2 \left( 1 + \frac{3M}{r} \right) \alpha^2 p + \rho h W^2 \left[ 2 \left(v_\theta v^\theta + v_\phi v^\phi \right) \right. \right. \nonumber \\
& &  \left. \left. - 2 \alpha^2 \left(1 + \frac{2 M^2}{r^2} + \frac{3 M}{r} \right) \left( v^r \right)^2 - \alpha v^r \right] \right\}.
\end{eqnarray}

The numerical procedure we follow in our code to solve the previous equations relies on a standard high-resolution shock-capturing scheme based on 
the Harteen, Lax, van Leer, Einfeldt (HLLE) Riemann solver \cite{hlle, einfeldt}. Moreover, we employ a monotonic, second-order (piecewise-linear) reconstruction of the rest-mass density $\rho$, the specific internal energy $\epsilon$, and three velocities (the {\it primitive} variables) in order to improve the spatial accuracy of our scheme. An important change with respect to the original version used in Ref.\ \cite{font_bondi_hoyle} is that the reconstruction procedure is applied not directly to the three velocities $v_r$, $v_\theta$, and $v_\phi$ but rather to the quantities $\sqrt{\gamma^{rr}}v_r = \mathrm{sign} (v_r) \sqrt{|v_r v^r|}$, $\sqrt{\gamma^{\theta \theta}}v_\theta = \mathrm{sign} (v_\theta) \sqrt{|v_\theta v^\theta|}$, $\sqrt{\gamma^{\phi \phi}}v_\phi = \mathrm{sign} (v_\phi) \sqrt{|v_\phi v^\phi|}$. (Note that no Einstein summation convention is used in the previous expressions.) The code is also second-order in time (third-order is implemented as well) as it uses a conservative, total-variation-diminishing Runge-Kutta scheme of second-order to perform the time update of the conserved quantities within a method-of-lines discretization of the differential equations \cite{shu_osher}. The recovery of the primitive quantities $\rho$, $\epsilon$, $v_r$, $v_\theta$, and $v_\phi$ is done through a Newton--Raphson root-finding algorithm.

We implement the boundary conditions in a standard way, using the ``ghost zones'' approach. More information on the particular choices of boundary conditions made for the models described in this paper can be found in Sec.\ \ref{sec_ic}.

\section{Code tests}
\label{sec_tests}

\subsection{Michel solution}

As a first consistency test we check that the new numerical code allows for a stable evolution of the so-called Michel solution~\cite{michel} (the relativistic analog of the Bondi solution~\cite{bondi}). This solution describes the spherically-symmetric, steady-state accretion of a perfect fluid on to a non-rotating black hole. We describe this test in detail because its implementation is used in the setup of our low angular momentum accretion model.

For a smooth, adiabatic flow the perfect gas equation of state $p = (\Gamma - 1)\rho \epsilon$ is equivalent to the polytropic relation $p = K \rho^\Gamma$, where $K$ is a constant. The specific enthalpy reads
\begin{eqnarray}
h =  \frac{\Gamma - 1}{\Gamma - 1 -\sos^2}\, , 
\end{eqnarray}
where $\sos$ is the local speed of sound. This quantity is related directly to the rest-mass density through the polytropic equation of state:
\begin{eqnarray}
\sos^2 = \frac{\Gamma p }{\rho h} = \frac{\Gamma - 1}{1 + \frac{\Gamma - 1}{K \Gamma} \rho^{1 - \Gamma}}\,. 
\end{eqnarray}

For the spherically symmetric, stationary, radial flow on the background metric (\ref{zzzc}), Eqs.~(\ref{zzza}) and (\ref{zzzb}) can be written as
\begin{eqnarray}
\label{yyya}
h \sqrt{1 - \frac{2M}{r} + (u^r)^2} = A, \;\;\; r^2 \rho u^r = B,
\end{eqnarray}
where $A$ and $B$ are integration constants. Here $u^\theta = u^\phi = 0$. Remarkably the above equations have precisely the same form also in polar coordinates (cf.\ Ref.\ \cite{mach_malec_karkowski}).

For the polytropic solutions that extend to infinity one can write the first of Eqs.~(\ref{yyya}) as
\begin{equation}
\label{newton_raphson_a}
\left( \Gamma - 1 - \sos^2 \right) = \sqrt{1 - \frac{2M}{r} + (u^r)^2} \left( \Gamma - 1 - \sosi^2 \right),
\end{equation}
where $\sosi$ is the asymptotic speed of sound. It was observed in \cite{sarbach, sarbach_mach} that for $\Gamma$ large enough (namely $\Gamma > 5/3$) there exist homoclinic-type solutions that are not defined globally; they are excluded from our setup.\footnote{ The term `homoclinic solutions' refers to the theory of dynamical systems. Introducing an auxilliary parameter (devoid of any physical interpretation), one can express the equations describing stationary Bondi-type (or Michel) flow as a dynamical system. Physical solutions correspond to the orbits of the system. It is quite surprising that for $\Gamma > 5/3$ there exist homoclinic orbits --- joining a critical point to itself \cite{sarbach_mach}.}

The value of the constant $B$ in Eqs.~(\ref{yyya}) can be chosen so that the solution passes through the sonic point -- a location on the diagram $u^r$ vs.\ $r$, where $(u^r/u_t)^2 = \sos^2$. The sonic point defined in this way is a relativistic analog of the Bondi sonic point, in which the radial velocity of the gas reaches the value of the speed of sound. The speed of sound at the sonic point is given by
\begin{eqnarray}
\soss^2 & = & \frac{1}{9} \Bigg\{ 6 \Gamma - 7 + 2 (3 \Gamma - 2) \cos \bigg[ \frac{\pi}{3} \nonumber \\
& & + \frac{1}{3} \mathrm{arc \, cos} \bigg\{ \frac{1}{2(3 \Gamma - 2)^3} \Big(54 \Gamma^3 - 351 \Gamma^2 + 558 \Gamma \nonumber \\
& & + 486 (\Gamma - 1) \sosi^2 - 243 \sosi^4 -259 \Big)  \bigg\}  \bigg] \Bigg\}.
\end{eqnarray}
The above formula is derived, e.g.\ in Ref.\ \cite{kinasiewicz_mach} (note that there is a misprint in the original formula in Ref.\ \cite{kinasiewicz_mach}). The location of the sonic point can be computed as
\begin{eqnarray}
r_\ast = \frac{M}{2} \frac{1 + 3 \soss^2}{\soss^2}\,, 
\end{eqnarray}
and the radial component of the four velocity of the fluid (in Eddington-Finkelstein coordinates) at the sonic point is
\begin{eqnarray}
u^r_\ast  = - \sqrt{\frac{\soss^2}{1 + 3 \soss^2}}\,.
\end{eqnarray}
Thus $B = \rho_\ast r_\ast^2 u_\ast^r$, where the rest-mass density at the sonic point $\rho_\ast$ can be, in turn, expressed in terms of $\soss$.

\begin{figure}[t!]
\begin{center}
\includegraphics[width=0.75\textwidth]{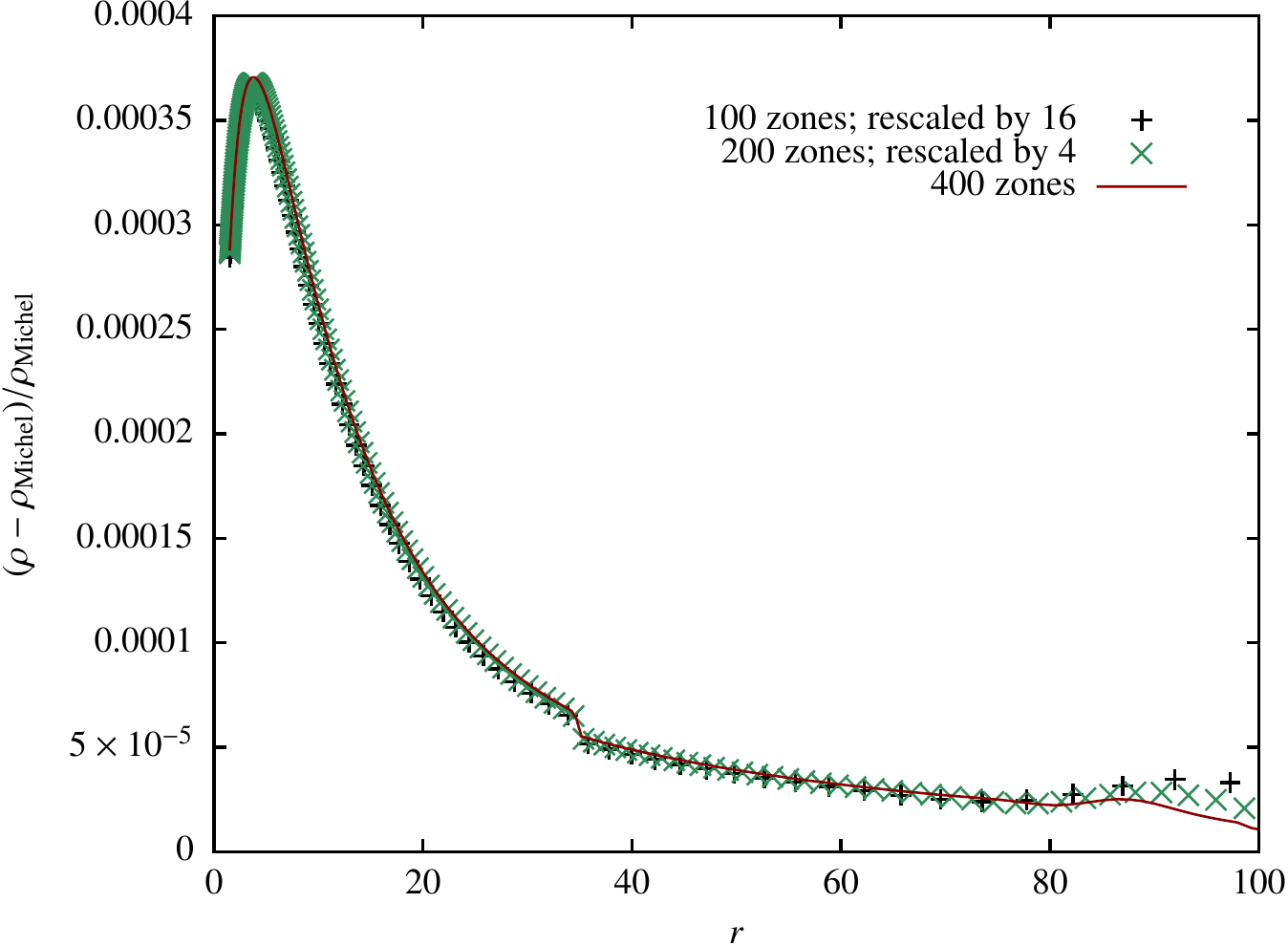}
\caption{\label{fig0a}A convergence test based on the Michel solution. The plot shows the relative error in the density $(\rho - \rho_\mathrm{Michel})/\rho_\mathrm{Michel}$ as a function of radius $r$. Here $\rho_\mathrm{Michel}$ denotes the (reference) rest-mass density of the Michel solution, and $\rho$ is the numerical value of the density after an evolution time $t = 100 M$. Different markings correspond to tests with 100, 200 and 400 zones spanning a range from $r_\mathrm{in} = 1.5 M$ to $r_\mathrm{out} = 100 M$. To demonstrate the 2nd order convergence, we rescale the errors corresponding to 100 radial zones by 16 and those corresponding to 200 zones by 4. The Michel solution is computed for $M=1$, $\sosi = 0.02236$, $\rho_\infty = 1$ and the polytropic index $\Gamma = 5/3$.There is a clear jump occurring at the sonic point radius $r = r_\ast \approx 34.8M$.}
\end{center}
\end{figure}

\begin{figure}[t!]
\begin{center}
\includegraphics[width=0.75\textwidth]{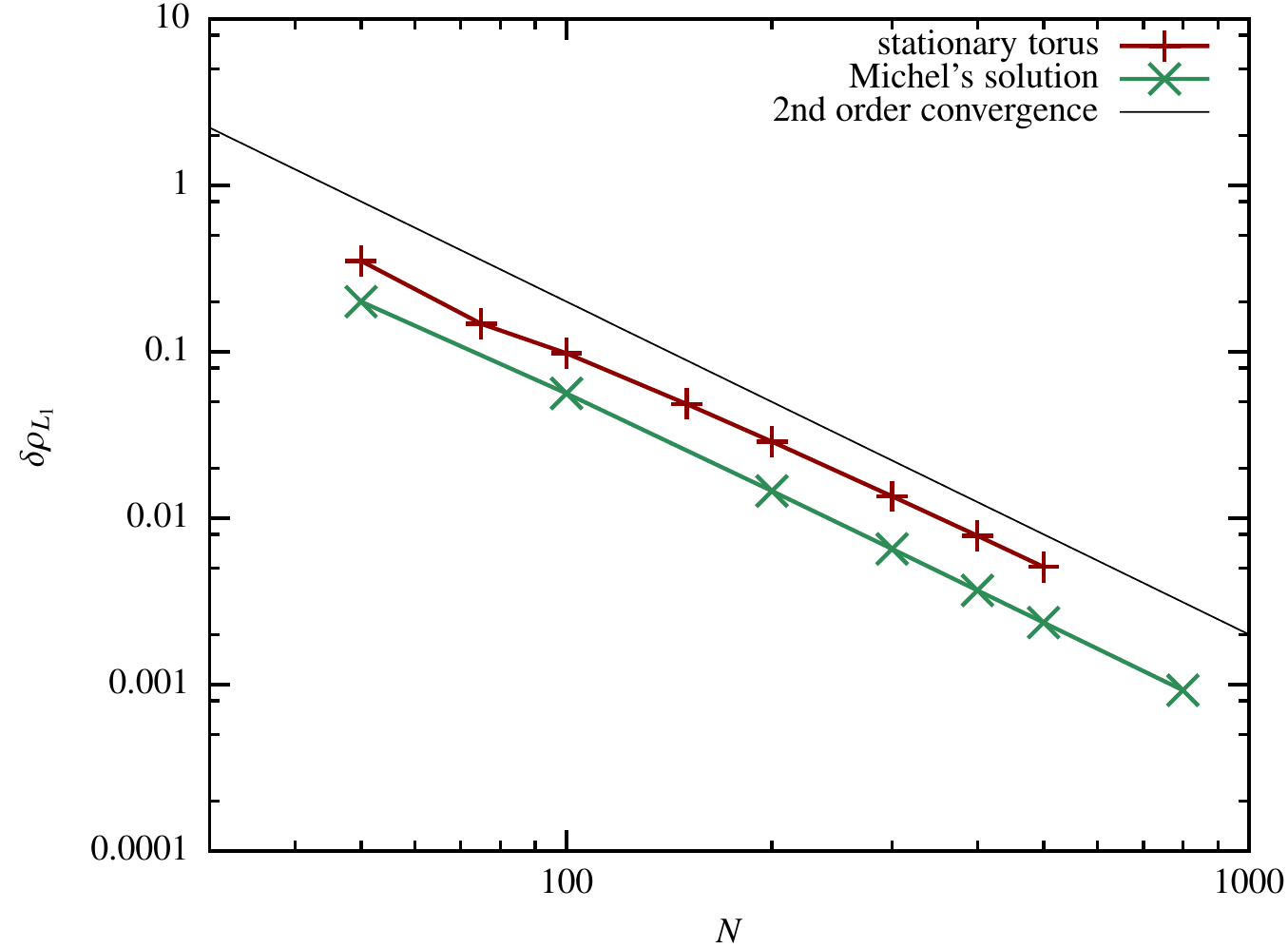}
\caption{\label{fig0c}Convergence tests based on the Michel (cross symbols) and Fishbone--Moncrief (plus symbols) solutions, as described in text. The thin solid line represents a function proportional to $N^{-2}$, depicting a perfect second order convergence.}
\end{center}
\end{figure}

In the next step, we find the values of $u^r$ from Eq.~(\ref{newton_raphson_a}) using a Newton--Raphson method. This part requires an explanation. Combining the second of Eqs.~(\ref{yyya}) with Eq.~(\ref{newton_raphson_a}) one can write
\begin{eqnarray}
1 - \frac{1}{1 + E \left[(u^r)^2\right]^\frac{\Gamma - 1}{2}} - \sqrt{1 - \frac{2M}{r} + (u^r)^2} \,C = 0\, , 
\label{eq28}
\end{eqnarray}
where
\begin{eqnarray}
E = \frac{\Gamma - 1	- \soss^2}{\soss^2} \left[ \frac{r^4}{r_\ast^4 (u^r_\ast)^2} \right]\,,  \;\;\;
C = 1 - \frac{\sosi^2}{\Gamma - 1}\,.
\end{eqnarray}

Eq.~(\ref{eq28}) can be solved for $(u^r)^2$. The local speed of sound is then computed according to
\begin{eqnarray} 
\sos^2 = \frac{\Gamma - 1}{1 + E (u^r)^{\Gamma - 1}}\,.
\end{eqnarray}
We also have
\begin{eqnarray}
\label{dddf}
 \epsilon = \frac{h - 1}{\Gamma} = \frac{\sos^2}{\Gamma (\Gamma - 1 - \sos^2)}, \;\;\; \rho = \left[ \frac{(\Gamma - 1) \epsilon}{K} \right]^\frac{1}{\Gamma - 1}.
\end{eqnarray}
The value of the polytropic constant $K$ can be controlled by assuming an asymptotic value of the rest-mass density $\rho_\infty$. Then
\begin{eqnarray}
K = \frac{\Gamma - 1}{\Gamma - 1 - \sosi^2} \, \frac{\sosi^2}{\Gamma \rho_\infty^{\Gamma - 1}}\,. 
\end{eqnarray}
In most cases we assume a normalized value $\rho_\infty = 1$. Note that this does not influence the dynamical properties of the Bondi--Michel accretion flow, i.e., the values of the speed of sound and the velocities of the gas are independent of this choice.

In the last step we obtain $v_r$ directly from $u^r$. The component $u^t$ (which is required to obtain $v_r$) can be computed from the normalization condition $g_{\mu \nu} u^\mu u^\nu = -1$. The final result reads
\begin{eqnarray}
v_r &=& \left( 1 + \frac{2M}{r} \right)^\frac{3}{2} \nonumber \\ &\times& \left\{ \left( 1 - \frac{2M}{r} \right) \frac{u^r}{ \frac{2M}{r} u^r + \sqrt{1 - \frac{2M}{r} + (u^r)^2}} + \frac{2M}{r + 2M} \right\} \,.
\end{eqnarray}

An example of the convergence test based on the stationary Michel solution is given in Fig.\ \ref{fig0a}. We evolve a transonic Michel solution corresponding to $M = 1$, $\sosi = 0.02236$, $\rho_\infty = 1$, and the polytropic index $\Gamma = 5/3$ on a numerical grid spanning from $r_\mathrm{in} = 1.5 M$ to $r_\mathrm{out} = 100 M$. The data in Fig.\ \ref{fig0a} correspond to different numbers of grid zones: 100, 200, and 400, respectively, and show the relative errors in the rest mass density after an evolution time $t = 100M$. The errors corresponding to 100 grid zones are divided by 16, while those corresponding to 200 grid points by 4. The overlapping of the graphs demonstrates the expected second-order convergence.

In addition, in Fig.\ \ref{fig0c} we plot with cross symbols the dependence of the (normalised) $L_1$ errors on the number of grid zones in the radial direction for the Michel solution for the same model parameters as in Fig.\ \ref{fig0a}. The $L_1$ norms are defined with respect to the induced metric at a given time slice. To be more specific, we define the errors as
\begin{eqnarray}
\label{l1}
\delta \rho_{L_1} \,  & = & \frac{1}{\mathrm{Vol} (V)} \int_V |\rho - \rho_0| \, \sqrt{\gamma} \,  dr \, d\theta \, d \phi \,, \\
\mathrm{Vol} (V) & =  &  \int_V \, \sqrt{\gamma} \,  dr \, d\theta \, d \phi \, ,
\label{volume}
\end{eqnarray} 
where $\rho_0$ denotes the exact Michel solution. In this case, the region $V$ is the spherical shell with the inner and outer radii equal to $r_\mathrm{in}$ and $r_\mathrm{out}$, respectively. The results shown in Fig.\ \ref{fig0c} indicate once again perfect second-order convergence of the code.

\subsection{Stationary torus}

We next test the code with evolutions of stationary torus (or thick disks).
Stationary test-fluid toroidal solutions in the Schwarzschild and Kerr metrics were constructed by Fishbone and Moncrief \cite{fishbone}. These solutions can be computed directly in Eddington--Finkelstein coordinates, but it turns out to be easier to obtain them using standard polar coordinates $(\hat t, \hat r, \hat \theta, \hat \phi)$ and then transform the angular velocity to the Eddington--Finkelstein form. We consider only the case of toroidal solutions around a non-rotating black hole. The Schwarzschild metric in polar coordinates reads
\begin{eqnarray}
 ds^2 & = & - \left( 1 - \frac{2M}{\hat r} \right) d \hat t^2 + \left( 1 - \frac{2M}{\hat r} \right)^{-1} d \hat r^2 \nonumber \\
& & + \hat r^2 \left( d \hat \theta^2 + \sin^2 \hat \theta d \hat \phi^2 \right).
\end{eqnarray}

We consider Fishbone--Moncrief disks with constant $j = u^{\hat t} u_{\hat \phi}$. The expression for the specific enthalpy is given by Eq.~(3.6) of Ref.\ \cite{fishbone}. For the Schwarzschild metric, with $\Delta = \hat r^2 - 2M \hat r$, $\Sigma = \hat r^2$ and $A = \hat r^4$ it yields
\begin{eqnarray}
\ln h & = & \frac{1}{2} \ln \left\{ \frac{1 + [1 + 4 j^2 \hat r^4 (\hat r^2 - 2 M \hat r)/(\hat r^4 \sin \hat \theta)^2]^{1/2}}{\hat r^2(\hat r^2 - 2M \hat r)/\hat r^4} \right\} \nonumber \\
& - & \frac{1}{2} \sqrt{1 + \frac{4 j^2 \hat r^4 (\hat r^2 - 2 M \hat r)}{(\hat r^4 \sin \hat \theta)^2}} - C 
\nonumber \\
& = & \frac{1}{2} \ln \left( \frac{1 + q}{1 - \frac{2M}{\hat r}} \right) - \frac{1}{2} q - C,
\label{eq_h}
\end{eqnarray}
where
\begin{eqnarray}
q = \sqrt{1 + \frac{4 j^2 (1 - 2M/\hat r)}{\hat r^2 \sin^2 \hat \theta}}\,.
\label{eq_q}
\end{eqnarray}
The integration constant $C$ can be computed by assuming that the inner rim of the disk is located at $\rin$. This gives
\begin{eqnarray}
C = \frac{1}{2} \ln \left( \frac{1 + q_\mathrm{in}}{1 - \frac{2M}{\rin}} \right) - \frac{1}{2} q_\mathrm{in}, \;\;\;
q_\mathrm{in} = \sqrt{1 + \frac{4 j^2 (1 - 2M/\rin)}{\rin^2}}.
\end{eqnarray}

The connection between the specific enthalpy $h$, the pressure $p$, the rest-mass density $\rho$ and the specific internal energy $\epsilon$ is straightforward. Inside the torus, the flow is smooth, and we can use the polytropic relations given by Eqs.~(\ref{dddf}) in the previous section.

The computation of the velocities is simple but subtle. The transformation between polar $(\hat t,\hat r, \hat \theta,\hat \phi)$ and Eddington--Finkelstein coordinates $(t, r, \theta, \phi)$ is given by
\begin{eqnarray}
dt = d\hat t  + \frac{1}{1 - \frac{2M}{\hat r}} \frac{2M}{\hat r} d\hat r, \;\;\; r = \hat r, \;\;\; \theta = \hat \theta, \;\;\; \phi = \hat \phi\,.
\end{eqnarray}
In the polar coordinates the motion is strictly azimuthal, with $u^{\hat r} = u^{\hat \theta} = 0$. Thus
\begin{eqnarray}
u^t = u^{\hat t} + \frac{1}{1 - \frac{2M}{\hat r}} \frac{2M}{\hat r} u^{\hat r} = u^{\hat t}\, , \;\;\; u^\phi = u^{\hat \phi}, \;\;\; u_\phi = u_{\hat \phi}\,.
\end{eqnarray}
The numerical code requires a prescription of
\begin{eqnarray} 
v_\phi = \frac{u_\phi}{\aef u^t} = \frac{u_{\hat \phi}}{\aef u^{\hat t}}, 
\end{eqnarray}
where $\aef = \sqrt{r/(r + 2M)}$ is the lapse function corresponding to Eddington--Finkelstein coordinates. Note that
\begin{eqnarray}
v_\phi = \frac{\alpha^2}{\aef} \frac{u^{\hat t} u_{\hat \phi}}{(\alpha u^{\hat t})^2} = \frac{\alpha^2}{\aef} \frac{j}{W^2} = \frac{\alpha^2}{\aef} j (1 - v_{\hat \phi} v^{\hat \phi}), 
\end{eqnarray}
where $\alpha = \sqrt{1 - 2M/r}$, and $W = 1/\sqrt{1 - v_{\hat \phi} v^{\hat \phi}}$ is the Lorentz factor in polar coordinates. The expression for $v_{\hat \phi}$ is analogous:
\begin{eqnarray}
v_{\hat \phi} = \alpha j (1 - v_{\hat \phi} v^{\hat \phi}) = \alpha j \left[ 1 - \frac{(v_{\hat \phi})^2}{r^2 \sin^2 \theta} \right]. 
\end{eqnarray}
It yields
\begin{eqnarray}
v_{\hat \phi} = \frac{(q - 1) r^2 \sin^2 \theta}{2 \alpha j}, 
\end{eqnarray}
where $q$ is given by Eq.~(\ref{eq_q}). Inserting the above expression into the formula for $v_\phi$, one obtains, after some calculations,
\begin{eqnarray}
v_\phi = \frac{(q - 1) r^2 \sin^2 \theta}{2 \aef j}. 
\end{eqnarray}
The full solution is specified by the following parameters: $j$, $\rin$, $K$, $\Gamma$. We consider only a part of the solution with $r > \rin$ (it is possible to obtain from Eq.\ (\ref{eq_h}) non-zero values of $h$ also for $r < \rin$).

\begin{figure}[t!]
\begin{center}
\includegraphics[width=0.75\textwidth]{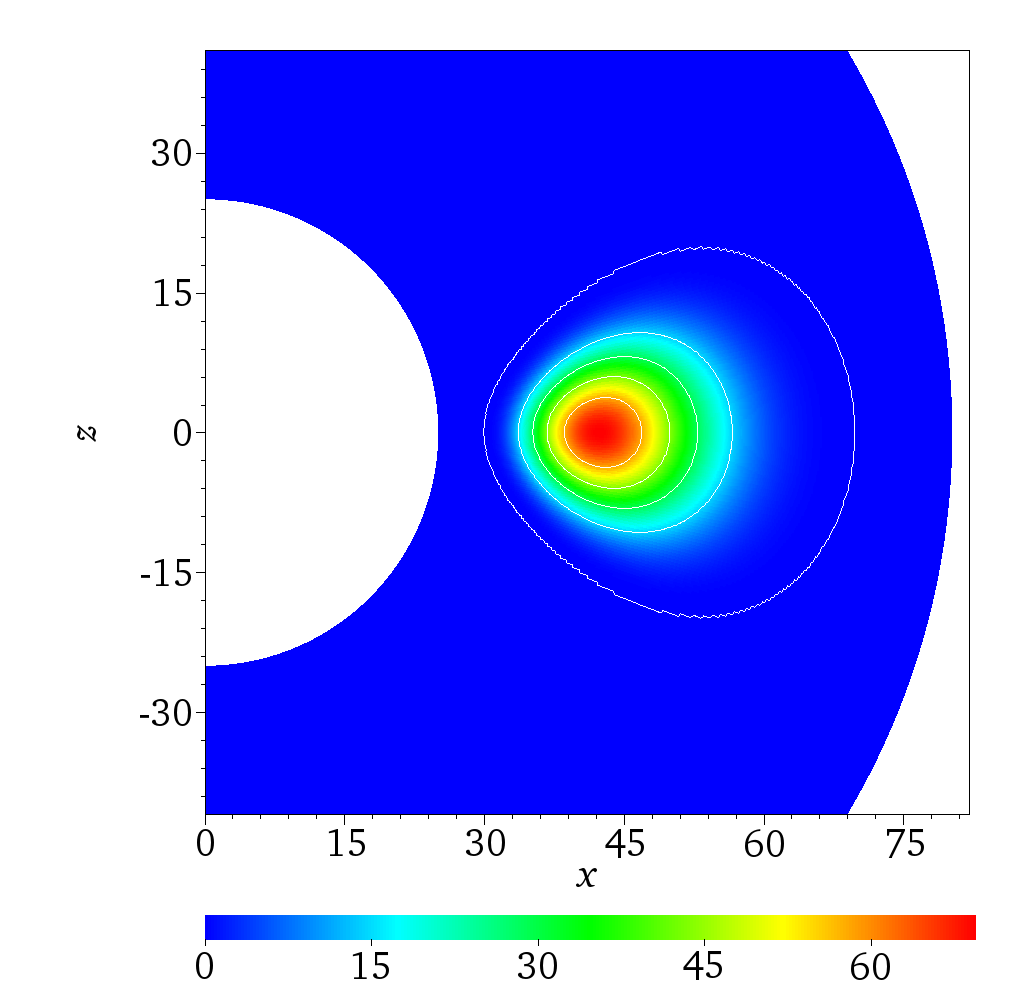}
\caption{\label{fig0b}The density of the Fishbone--Moncrief disk used as a test solution. The white lines in the plot denote the surfaces of constant density. The outermost line corresponds to the boundary of the disk. We assume units with $G = c = M = 1$.}
\end{center}
\end{figure}

Because grid-based numerical codes cannot handle a vacuum region outside of the disk, we follow the ``artificial atmosphere'' approach. The values of the hydrodynamical quantities within the numerical grid are restored to the artificial atmosphere levels whenever the rest-mass density $\rho$ or the azimuthal velocity $v_\phi$ are small enough. 

The test based on the above stationary solution is similar to the one performed in Ref.\ \cite{harm1}, and it is also similar to the previous test making use of the Michel solution. We evolve initial data consisting of a stationary disk and the artificial atmosphere up to time $t = 100 M$\footnote{This is a short time compared to the orbital period of the disk, but it is long enough to perform the convergence test. The orbital period for circular motion in a constant-$j$ disk can be computed as
\[ t_{\rm orb} = \frac{\pi r^2 \left[ 1 + \sqrt{1 + \frac{4 j^2}{r^2} \left(1 - \frac{2M}{r} \right) } \right] }{j \left( 1 - \frac{2M}{r} \right)} \]
at the equator. For $M=1$, $j=7$, and $r = 30$ the above formula yields $t_{\rm orb} \approx 900$.}, and check for the numerical error that appears in the solution. The error of the rest-mass density is measured with the (normalized) $L_1$ norm defined by Eq.\ (\ref{l1}), where $V$ denotes the volume of the torus, and $\rho_0$ is the analytic, stationary solution. As an example, we consider a stationary solution with the following parameters: $M = 1$, $j = 7$, $\rin = 30$, $\Gamma = 4/3$. The rest-mass density within such a disk is plotted in Fig.\ \ref{fig0b}. Its maximum value at the center of the torus is of the order of $70$. To minimize the influence of the artificial atmosphere we restrict the region $V$ in which the error is computed to the part of the disk where $\rho \ge 1$. The results of the convergence test are shown using plus symbols in Fig.\ \ref{fig0c}. The errors are computed for a handful of numerical grids of resolution $N \times N$ each. As the figure shows we obtain second order convergence -- this can be easily seen by comparing the obtained errors with the plot of a function $\sim N^{-2}$. 

\begin{table}
\begin{center}
\begin{tabular}{ccccccccc}
\hline
\hline
No. & $r^\prime_\mathrm{S}$ &  $\sosi$ & $r_\mathrm{B}$ & $r_\mathrm{out}$ & $l_0$ & $r_\mathrm{C}$ & $r_\mathrm{C}^\prime$ & $v^\theta$ \\
\hline
1a & $10^{-2}$   & 0.07071 & 200  & 300  & 6.657  & 40   & $2 \times 10^{-1}$   &  free \\
2a & $10^{-3}$   & 0.02236 & 2000 & 3000 & 4.5714 & 16   & $8 \times 10^{-3}$   &  free \\
3a & $10^{-3}$   & 0.02236 & 2000 & 3000 & 10.204 & 100  & $5 \times 10^{-2}$   &  free \\
4a & $10^{-3.5}$ & 0.01257 & 6324 & 9487 & 14.285 & 200  & $3.1 \times 10^{-2}$ &  free \\
\hline
1b & $10^{-2}$   & 0.07071 & 200  & 300  & 6.657  & 40   & $2 \times 10^{-1}$   &  fixed \\
2b & $10^{-3}$   & 0.02236 & 2000 & 3000 & 4.5714 & 16   & $8 \times 10^{-3}$   &  fixed \\
3b & $10^{-3}$   & 0.02236 & 2000 & 3000 & 10.204 & 100  & $5 \times 10^{-2}$   &  fixed \\
4b & $10^{-3.5}$ & 0.01257 & 6324 & 9487 & 14.285 & 200  & $3.1 \times 10^{-2}$ &  fixed \\
\hline
\hline
\end{tabular}
\caption{\label{table1} Parameters of the models discussed in this paper. From left to right the columns report: the name of the model, the ratio of the Schwarzschild radius to the Bondi radius, $r^\prime_\mathrm{S}$, the asymptotic speed of sound, $\sosi$, the Bondi radius, $r_\mathrm{B}$, the outer radius of the grid, $r_\mathrm{out}$, a constant to introduce rotation to the Michel solution, $l_0$, the circularization radius, $r_\mathrm{C}$, both as such and as  normalized by the Bondi radius, $r_\mathrm{C}^\prime$, and the outer boundary conditions for the polar velocity $v^\theta$. We assume $G=c=M=1$.}
\end{center}
\end{table}

\section{Initial data, boundary conditions and accretion rates}
\label{sec_ic}

We construct our initial models of low angular momentum accretion using the Michel solution as a background one. In Ref.\ \cite{proga_begelman} Proga and Begelman used the classical Bondi-type solution as a background solution computed for the pseudo-Newtonian gravitational potential of Ref.\ \cite{paczynski_wiita}. In order to keep a connection with the results of~\cite{proga_begelman} we follow the same notation to parametrize the solutions. This notation is based on two characteristic scales that appear in the Michel accretion model: the Schwarzschild radius $r_\mathrm{S} = 2M$, and the Bondi radius defined as $r_\mathrm{B} = M \sosi^{-2}$. Given the asymptotic speed of sound $\sosi$ we can also compute the ratio $r_\mathrm{S}^\prime = r_\mathrm{S}/r_\mathrm{B} = 2 \sosi^2$. Our sample of initial models are reported in Table \ref{table1}. We discuss next how those models are built.

\subsection{Adding rotation to the initial data}

Following Proga and Begelman \cite{proga_begelman} we add rotation to the initial data so that the distribution of the specific angular momentum has the form $l = l_0 f(\theta)$ for $r > \delta r_\ast$, where $r_\ast$ is the sonic radius, $l_0$ is a constant, and $\delta$ is a parameter of order unity. Our main choice for $f(\theta)$ is $f(\theta) = 1 - |\cos \theta|$. This admittedly artificial form of $f(\theta)$ was also used in Ref.\ \cite{proga_begelman}.

In the relativistic case one has to decide on a definition of specific angular momentum. Different possibilities are discussed e.g.\ in Ref.\ \cite{kozlowski}. Here we assume the definition $l = - u_\phi/u_t$, so that $l = -W v_\phi/u_t$, and 
\begin{eqnarray}
u_t & = & g_{tt} \frac{W}{\alpha} + \beta_r u^r = g_{tt} \frac{W}{\alpha} + \beta_r W \left( v^2 - \frac{\beta^r}{\alpha} \right) \nonumber = -\alpha W + W \beta_r v^r.
\end{eqnarray}
Thus
\begin{eqnarray} 
l = \frac{v_\phi}{\alpha - \beta_r v^r} \,,
\end{eqnarray}
or
\begin{eqnarray}
\label{llla}
 v_\phi = l (\alpha - \beta_r v^r).
\end{eqnarray}
The last equation can be used to introduce rotation to the original background Michel solution. We set
\begin{equation}
\label{llac}
v_\phi = l_0 f(\theta) \left( \alpha - \beta_r v^r \right), \quad r > \delta r_\ast,
\end{equation}
as an initial condition.

The constant $l_0$ can be specified using the notion of the circularization radius $r_\mathrm{C}$, i.e., the radius of the orbit of the test particle with a given specific angular momentum (or velocity). This notion has a clear meaning in Newtonian theory. In general relativity it depends on the definition of the specific angular momentum. In our case the appropriate relation can be obtained as follows. Assume that $u^r = u^\theta = 0$ (this means that $v^r \neq 0$). The geodesic equation reads
\begin{eqnarray} 
\Gamma^\mu_{\nu \rho} u^\nu u^\rho = 0. 
\end{eqnarray}
Assuming an equatorial motion ($\theta = \pi/2$), one obtains
\begin{eqnarray}
\frac{M}{r_\mathrm{C}^3}(u^t)^2 - (u^\phi)^2 = 0. 
\end{eqnarray}
For the specific angular momentum defined as $l = - u_\phi/u_t$ this gives the relation
\begin{equation}
\label{laab}
l^2 = \frac{M r_\mathrm{C}}{\left(1 - \frac{2M}{r_\mathrm{C}} \right)^2},
\end{equation}
which generalizes the Newtonian formula $l^2 = GMr_\mathrm{C}$. 

In summary, the initial data are constructed as follows. The density $\rho$, the specific internal energy $\epsilon$, and the radial component of the three-velocity $v_r$ are given by the stationary Michel solution. The azimuthal component of the three-velocity $v_\phi$ is computed from Eq.\ (\ref{llac}), and the value of $l_0$ is determined by assuming a corresponding circularization radius, as in Eq.\ (\ref{laab}). We set the meridional velocity component $v_\theta$ to zero everywhere (actually $v_\theta = 0$ for Michel's solution as well). The latter assumption is actually incompatible with the dynamics of the flows described in this paper. The velocity component $v_\theta$ is set to zero only initially, but it is non-zero during the subsequent evolutions of all investigated models.

\subsection{Boundary conditions}

The implementation of the boundary conditions for our problem is a subtle issue. In this paper we generally follow the scheme used in Ref.\ \cite{proga_begelman}, making slight adjustments suitable for the relativistic formulation.

We use a spherical grid with the radial variable ranging from $r_\mathrm{in}$ to $r_\mathrm{out}$. The grid is logarithmic in the radial direction and equidistant in the angular ones. At the inner radial boundary of the grid (at $r_\mathrm{in}$) we impose outflowing conditions, appropriate for the matter flowing freely into the black hole -- the values in the ghost zones are either extrapolated from the adjacent zones of the grid (we use a linear extrapolation), or simply copied from the first zone of the grid. In any case the inner boundary is always located beneath the horizon of the black hole, i.e., $r_\mathrm{in} < 2M$. Usually we set $r_\mathrm{in} = 1.5 M$. In this way one ensures that a possible undesirable reflection from the boundary of the numerical grid does not influence the region outside the black hole. This is one of the key features that distinguishes a relativistic simulation from its Newtonian counterpart, and it comes as a great advantage when using horizon-penetrating coordinates such as Eddington-Finkelstein.

At the outer boundary of the grid we use a mixture of free (outflowing) and fixed boundary conditions. The values of the rest-mass density $\rho$ and the specific internal energy $\epsilon$ are kept fixed as given by Michel's solution. The radial velocity is free, i.e.~the values in the ghost zones are copied from the outermost zone of the grid. The component $v_\theta$ is either kept fixed ($v_\theta = 0$) or evolved freely, similarly to the radial component of the velocity. The azimuthal component is set ``by hand'' in such a way that the specific angular momentum is prescribed at the boundary by the same formulae that are used for the initial data. This procedure is slightly different from simply setting $v_\phi$ at the a priori prescribed values because the relation between $v_\phi$ and $l$ [Eq.\ (\ref{llla})] involves the radial velocity component, which is evolved freely. We have also experimented with outflowing conditions at the outer boundary of the grid.

Other boundary conditions, at the axes and in the direction of the $\phi$ coordinate, are standard and follow from the symmetry requirements.

\subsection{Accretion rates}

Accretion rates are computed according to definitions used in Ref.\ \cite{font}. For the mass accretion rate we set
\begin{eqnarray}
\dot m & = & \int_0^\pi d \theta \int_0^{2\pi} d\phi \frac{\partial^2 \dot m}{\partial \phi \partial \theta}, \\
 \frac{\partial^2 \dot m}{\partial \phi \partial \theta} & = & \sqrt{-g} \rho u^r = r^2 \sin\theta D \left( v^r - \frac{\beta^r}{\alpha} \right).
\end{eqnarray}
The angular momentum accretion rate $\dot L$ is computed assuming $l = - u_\phi/u_t$ as our definition for the specific angular momentum
\begin{eqnarray}
\label{ggh5}
\dot L & = & \int_0^\pi d\theta \int_0^{2\pi} d\phi \frac{\partial^2 \dot m}{\partial \phi \partial \theta} l 
 \\
& = & r^2 \int_0^\pi d \theta \int_0^{2\pi} d\phi  \sin \theta D \left( v^r - \frac{\beta^r}{\alpha} \right) \frac{v_\phi}{\alpha - \beta^r v_r}. \nonumber
\end{eqnarray}
In axial symmetry, which we adopt in the numerical simulations discussed in this work, the above formulae simply read
\begin{eqnarray}
\dot m & = & \int_0^\pi d \theta \frac{d \dot m}{d \theta}, \\
 \frac{d \dot m}{d \theta} & = & 2 \pi \sqrt{-g} \rho u^r = 2 \pi r^2 \sin \theta D \left( v^r - \frac{\beta^r}{\alpha} \right)
\end{eqnarray}
and
\begin{eqnarray}
\label{ggh6}
\dot L & = & \int_0^\pi d\theta \frac{d \dot m}{d \theta} l = 2 \pi r^2 \int_0^\pi d \theta \sin \theta D \left( v^r - \frac{\beta^r}{\alpha} \right) \frac{v_\phi}{\alpha - \beta^r v_r}.
\end{eqnarray}
These quantities are computed at the horizon ($r = 2M$). We evaluate the above integrals using standard quadrature formulae.

The reference value for $\dot m$ is given by the accretion rate of the unperturbed Michel flow, $\dot m_0$. It can be computed as
\begin{eqnarray}
\dot m_0 & = & - 4 \pi r_\ast^2 \rho_\ast u^r_\ast = \pi M^2 \left( \frac{1 + 3 \soss^2}{\soss^2} \right)^\frac{3}{2} \left( \frac{\soss^2}{\sosi^2} \frac{\Gamma - 1 - \sosi^2}{\Gamma - 1 - \soss^2} \right)^\frac{1}{\Gamma - 1} \rho_\infty. 
\end{eqnarray}

\section{Numerical results}
\label{sec_results}

\begin{figure}[t!]
\begin{center}
\includegraphics[width=0.75\textwidth]{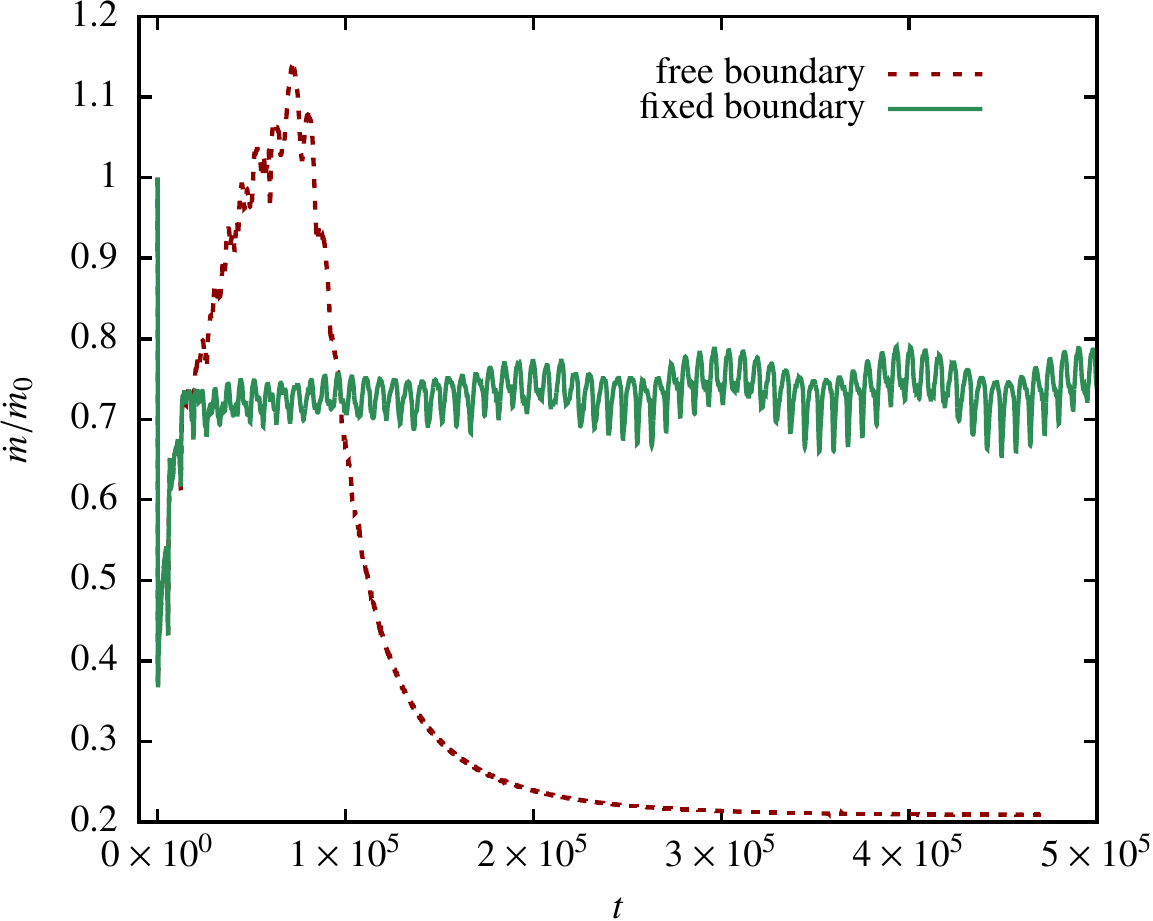}
\caption{\label{m20}Time evolution of the mass accretion rate for models 1a (``free $v^\theta$'' at the boundary; red dashed line) and 1b (``fixed $v^\theta$'' at the boundary; solid green line) from Table \ref{table1}.}
\end{center}
\end{figure}

\begin{figure}[t!]
\begin{center}
\includegraphics[width=0.75\textwidth]{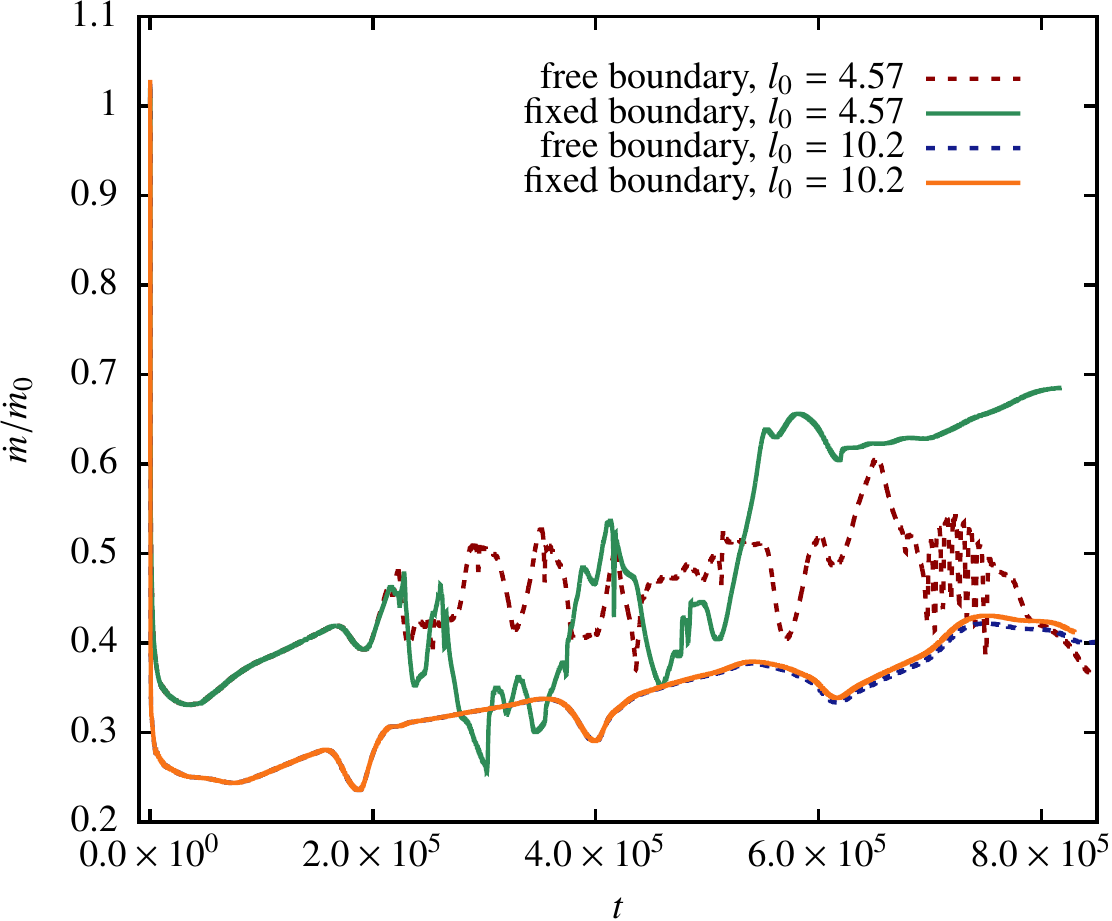}
\caption{\label{m30}Time evolution of the mass accretion rate for models 2a (red dashed curve), 2b (green solid curve), 3a (black dashed curve), and 3b (orange solid curve).}
\end{center}
\end{figure}

\begin{figure}[t!]
\begin{center}
\includegraphics[width=0.75\textwidth]{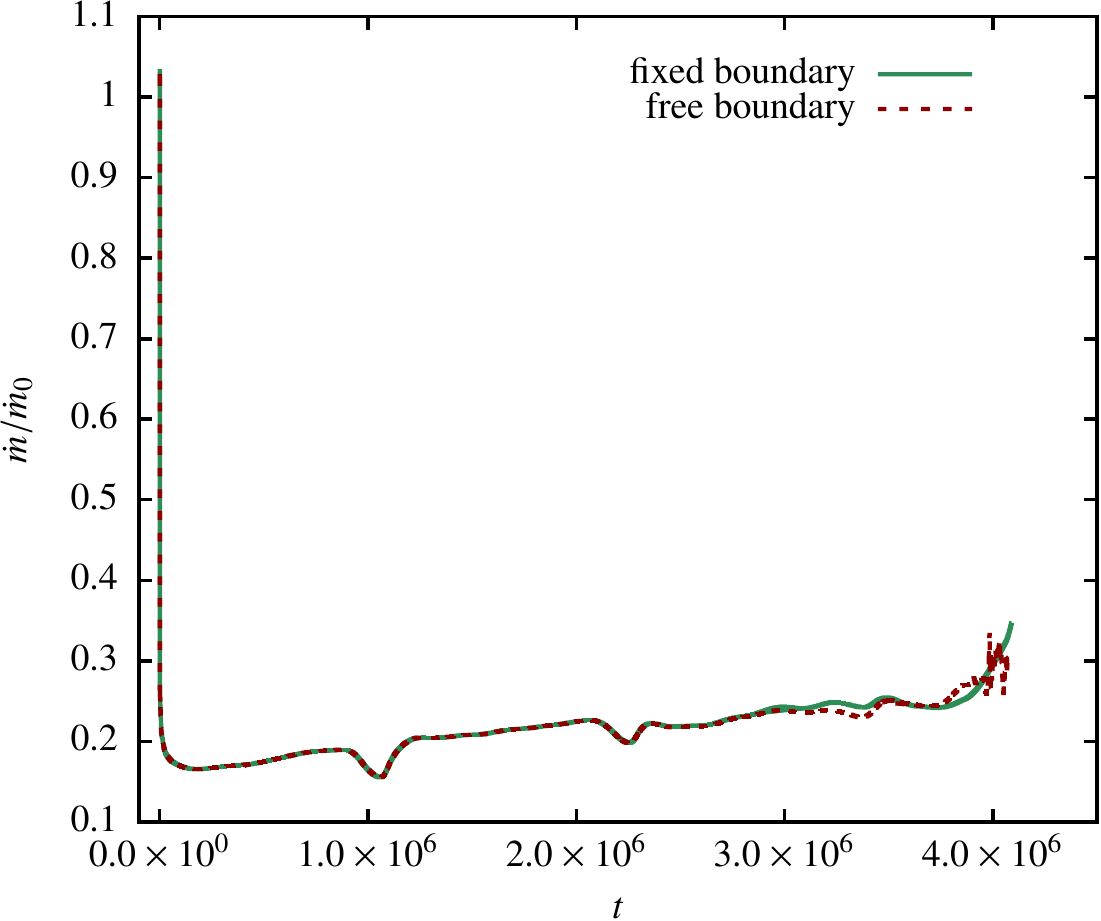}
\caption{\label{m35}Time evolution of the mass accretion rate for models 4a (``free $v^\theta$'' at the boundary) and 4b (``fixed $v^\theta = 0$'' at the boundary).}
\end{center}
\end{figure}

\begin{figure}[t!]
\begin{center}
\includegraphics[width=0.75\textwidth]{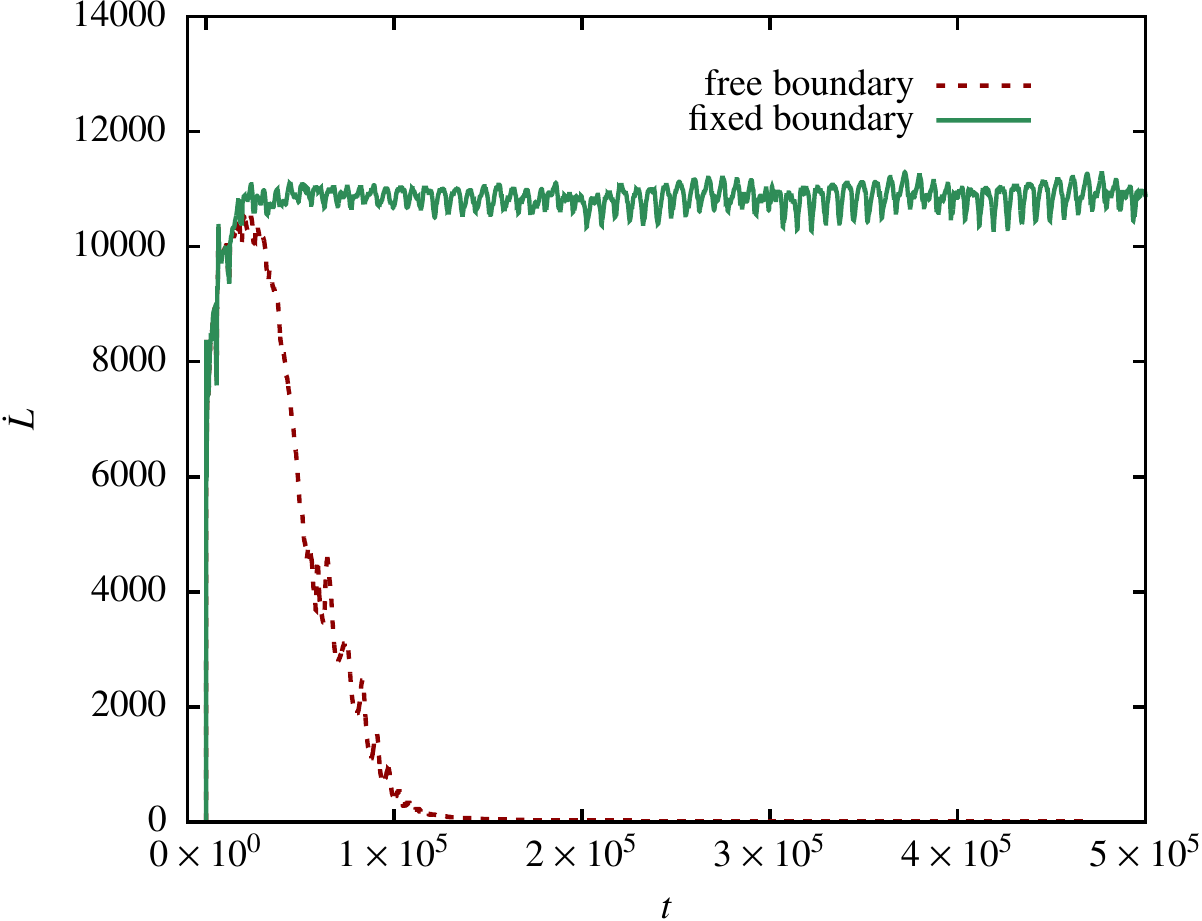}
\caption{\label{l20}Time evolution of the angular momentum accretion rate for models 1a (``free $v^\theta$'' at the boundary; red dashed line) and 1b (``fixed $v^\theta = 0$'' at the boundary; solid green line).}
\end{center}
\end{figure}

We compute a total of eight accretion models for different parameters characterizing the initial data, and different boundary conditions,  both in 2D (assuming axial symmetry) and in 3D. Here, for the sake of comparison with the results of~\cite{proga_begelman} we only discuss the results of our axially symmetric simulations. The results of 3D models will be discussed elsewhere. The parameters describing the 2D models are collected in Table \ref{table1}. The eight models can be grouped in two subsets of four models each. Models labelled ``a'' or ``b'' in either subset differ only on the outer boundary condition for the polar velocity $v^{\theta}$, free in the former and fixed in the latter.

The behavior of the accretion flow is mainly influenced by the value of the asymptotic speed of sound $\sosi$. Following Ref.\ \cite{proga_begelman} we prescribe $\sosi$ by setting the value of the ratio of the Schwarzschild and Bondi radii $r^\prime_\mathrm{S} = r_\mathrm{S}/r_\mathrm{B} = 2\sosi^2$. We choose in particular the values $r^\prime_\mathrm{S} = 10^{-2}, 10^{-3}$, and $10^{-3.5}$. This corresponds to $\sosi = 0.07071, 0.02236$, and $0.01257$, respectively. In all cases the inner radial boundary of the grid is located within the event horizon, at $r_\mathrm{in} = 1.5M$. The outer boundary of the grid is placed at $r_\mathrm{out} = 1.5 r_\mathrm{B}$. Thus, the value of $r_\mathrm{out}$ varies between $300 M$ (for $\sosi = 0.07071$) and $9500 M$ (for $\sosi = 0.01257$). 

The simulations that we discuss here were performed assuming $\Gamma = 5/3$, and normalizing conditions $M = 1$ and $\rho_\infty = 1$. The assumed form of the initial distribution of the angular momentum is given by choosing $f(\theta) = 1 - |\cos \theta|$ (cf. Sec.\ \ref{sec_ic}). We have also experimented with other choices but this one seems to be representative of the type of fluid evolutions encountered. The values of the parameter $l_0$ and corresponding parameters $r_\mathrm{C}$ and $r^\prime_\mathrm{C} = r_\mathrm{C}/r_\mathrm{B}$ are listed in Table \ref{table1}. As mentioned before, we consider two different implementations of the outer radial boundary conditions, as described in Sec. \ref{sec_ic}; they differ by the way in which the meridional component of the velocity $v^\theta$ is treated. In the ``b'' models that are denoted in Table \ref{table1} as $v^\theta$ fixed, $v^\theta$ is set to zero at the outer boundary. Correspondingly, in the ``a'' models denoted with ``$v^\theta$ free'', it is evolved using outflowing boundary conditions -- the value of $\mathrm{sgn} (v_\theta) \sqrt{|v_\theta v^\theta|}$ from the outermost zones is copied to the ghost zones.

Our simulations show that the dynamical evolution of the models is very sensitive to the value of the asymptotic speed of sound $\sosi$. The general trend is that for low (more realistic) values of the asymptotic speed of sound, the addition of angular momentum in the initially symmetric distribution of matter results in the development of a ``turbulent'' pattern in the specific angular momentum, eventually breaking the symmetry. This does not happen for models with high values of the asymptotic speed of sound. Moreover, the accretion flow is characterized by the appearance of a thick torus in the equatorial plane. This torus is built from material with sufficiently high angular momentum to slow down accretion. By comparing the models with our three different choices of $\sosi$, we also find that the dependence of the behavior of the model on the choice of outer boundary conditions becomes more apparent the larger $\sosi$ becomes. For the smallest, more realistic value of our sample, $\sosi = 0.01257$, the evolutions show almost no dependence on the boundary conditions, giving more credence to the obtained results. All of these findings are discussed next by providing figures showing the time evolution of the specific angular momentum and the rest-mass density, along with the accretion rates of mass and angular momentum.

\begin{figure*}
\begin{center}
\includegraphics[width=\textwidth]{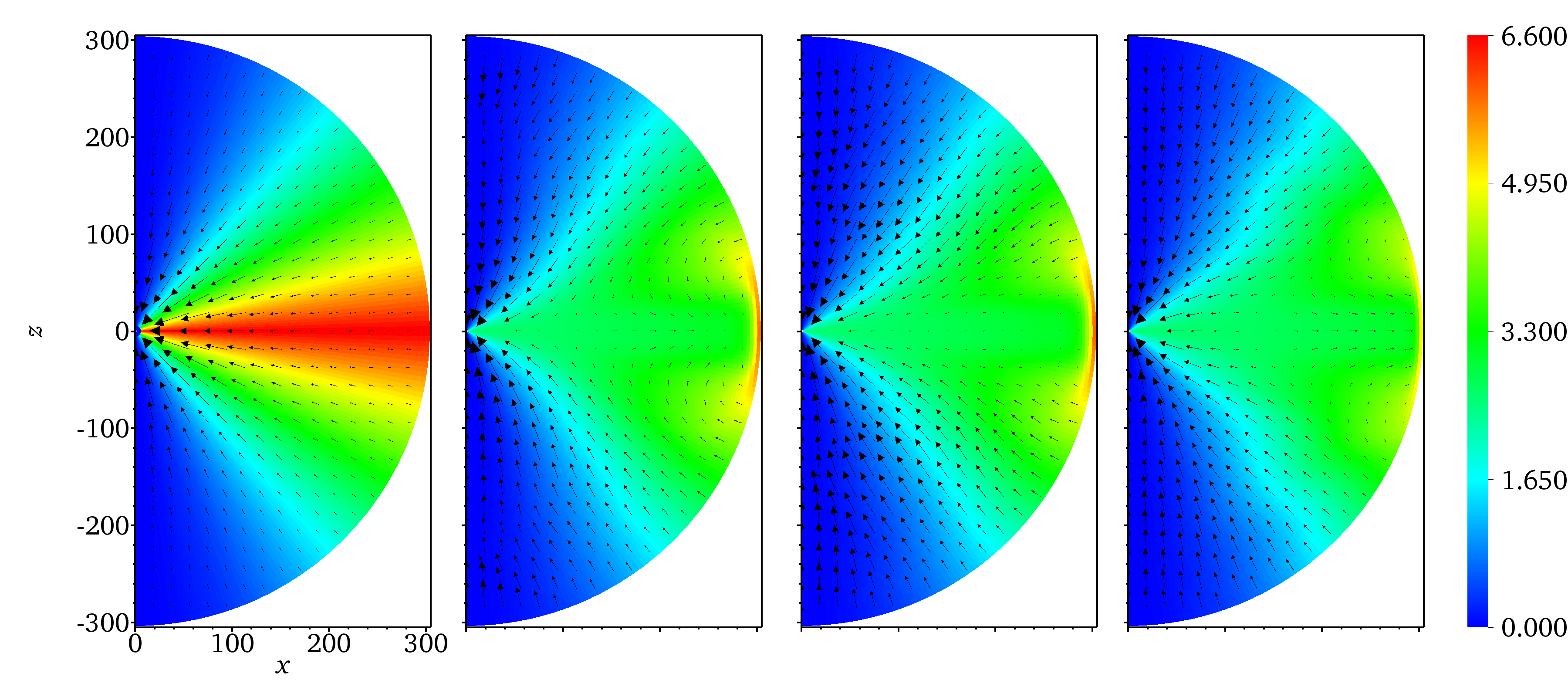}
\caption{\label{20_ang}Model 1b from Table \ref{table1}. The plots show, from left to right, a sequence in the evolution of the angular momentum (color coded) and of the velocity in the meridional plane (arrows) at times $t = 0$, $12.5\times 10^4$, $25\times 10^4$, and $37.5 \times 10^4$.}
\end{center}
\end{figure*}

\begin{figure*}
\begin{center}
\includegraphics[width=0.8\textwidth]{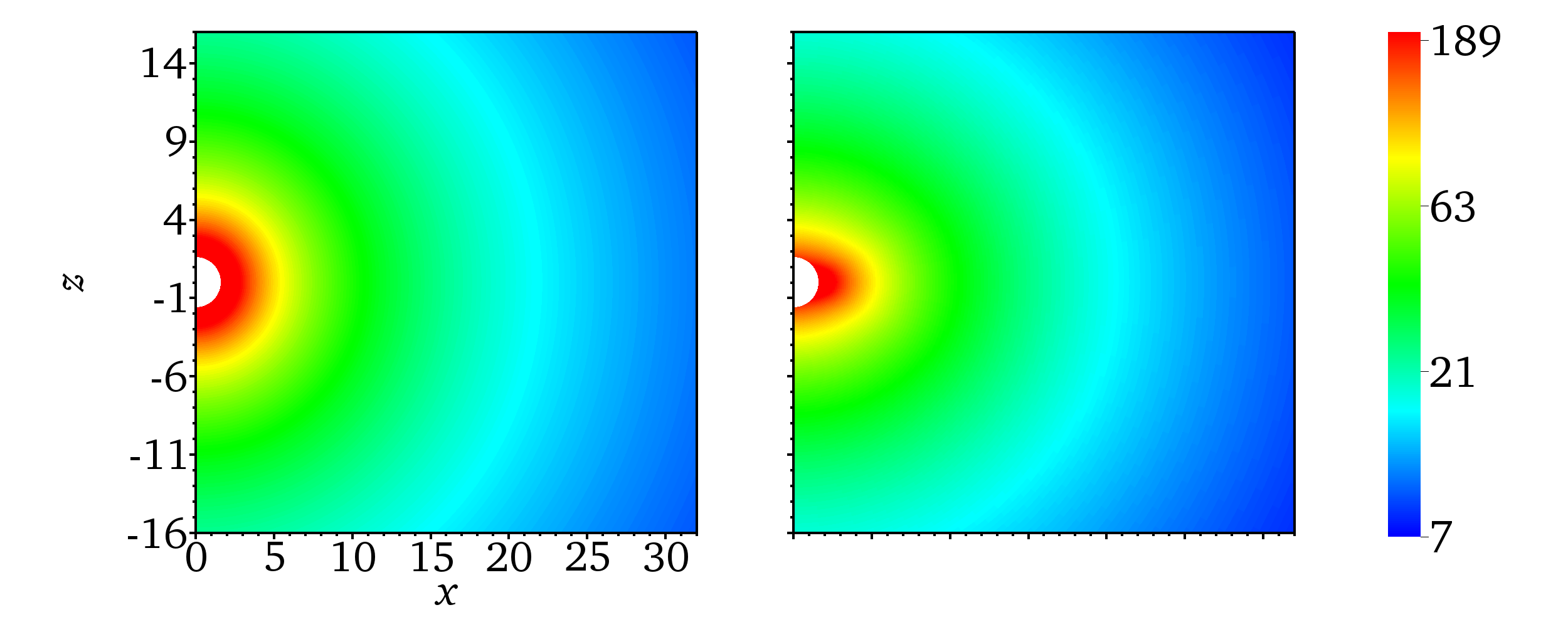}
\caption{\label{20_dens}Model 1b from Table \ref{table1}. The plots show, from left to right, the gas density $\rho$ at the initial time $t = 0$, and at the end of the evolution, $t=37.5 \times 10^4$. The asymptotic value of the density is $\rho_\infty = 1$.}
\end{center}
\end{figure*}

\begin{figure*}
\begin{center}
\includegraphics[width=\textwidth]{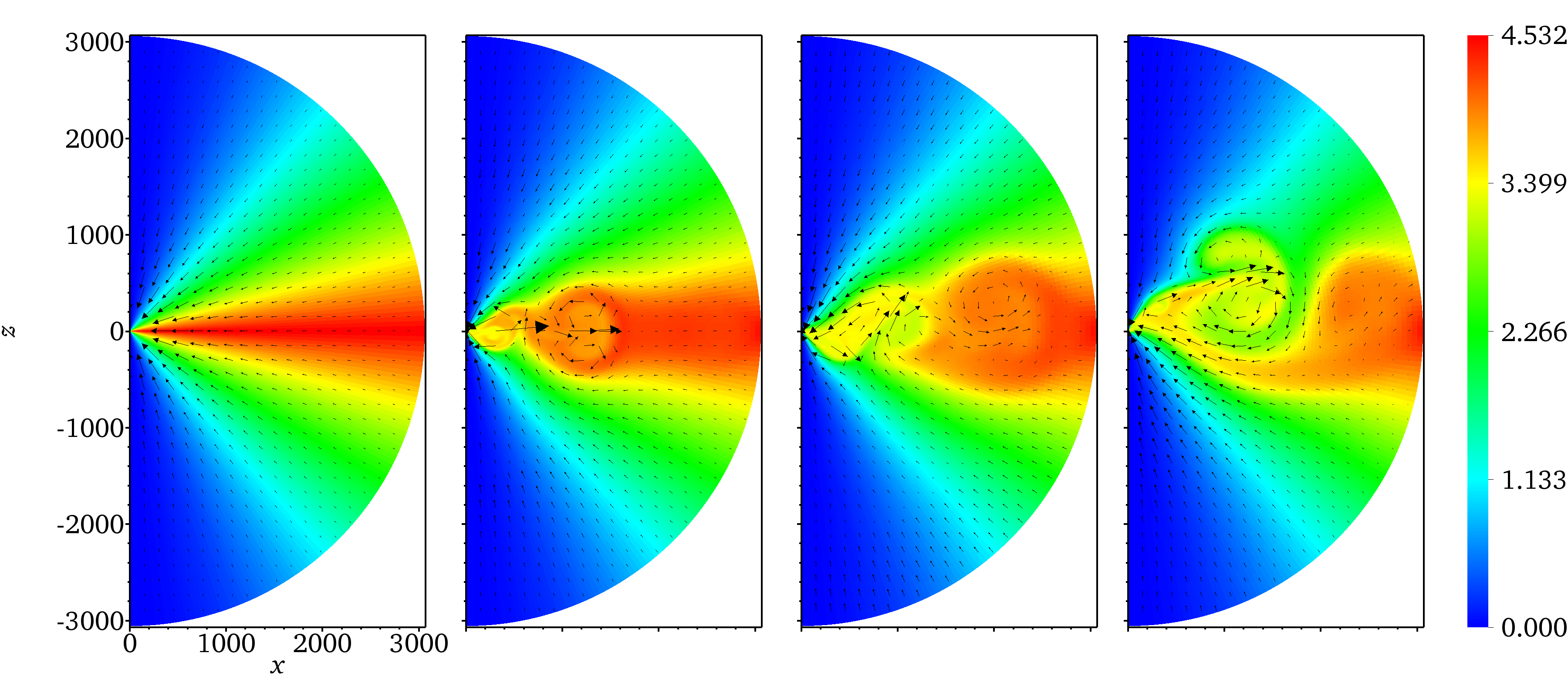}
\caption{\label{30_ang_free}Model 2a from Table \ref{table1}. The plots show, from left to right, a sequence in the evolution of the angular momentum (color coded) and of the velocity in the meridional plane (arrows) at times $t = 0$, $25\times 10^5$, $50\times 10^5$, and  $75 \times 10^5$.}
\end{center}
\end{figure*}

\begin{figure*}
\begin{center}
\includegraphics[width=\textwidth]{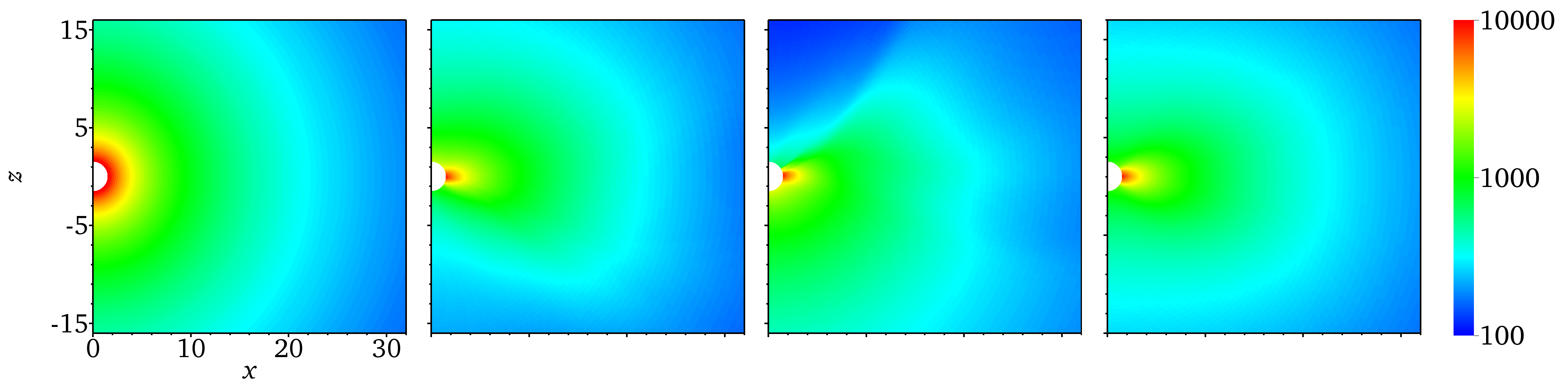}
\caption{\label{30_dens_free}Model 2a from Table \ref{table1}. The plots show, from left to right, a sequence in the evolution of the density $\rho$ at times $t = 0$, $25\times 10^5$, $50\times 10^5$, and  $75 \times 10^5$. The asymptotic value of the density is $\rho_\infty = 1$.}
\end{center}
\end{figure*}

\begin{figure*}
\begin{center}
\includegraphics[width=\textwidth]{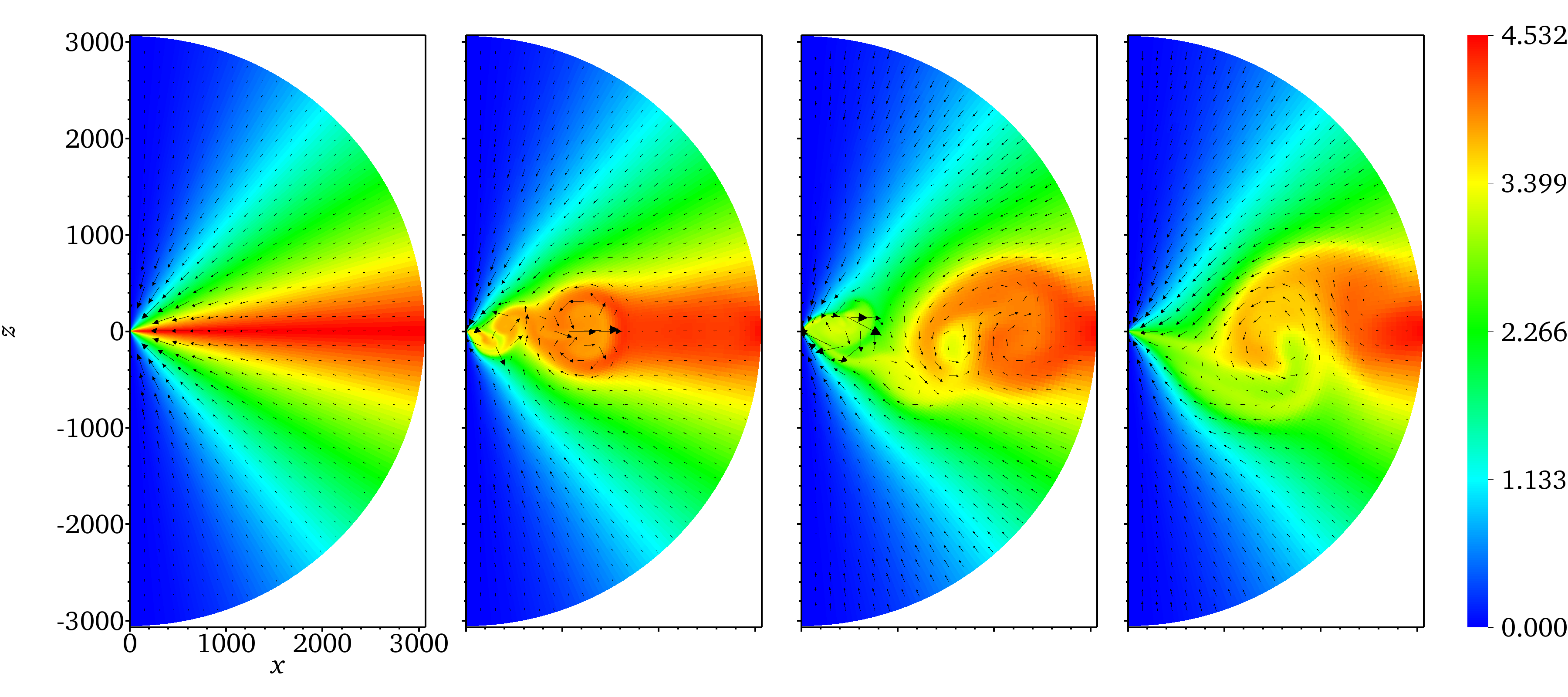}
\caption{\label{30_ang_fixed}Model 2b from Table \ref{table1}. The plots show, from left to right, a sequence in the evolution of the angular momentum (color coded) and of the velocity in the meridional plane (arrows) at times $t = 0$, $25\times 10^5$, $50\times 10^5$, and  $75 \times 10^5$.}
\end{center}
\end{figure*}

\begin{figure*}
\begin{center}
\includegraphics[width=\textwidth]{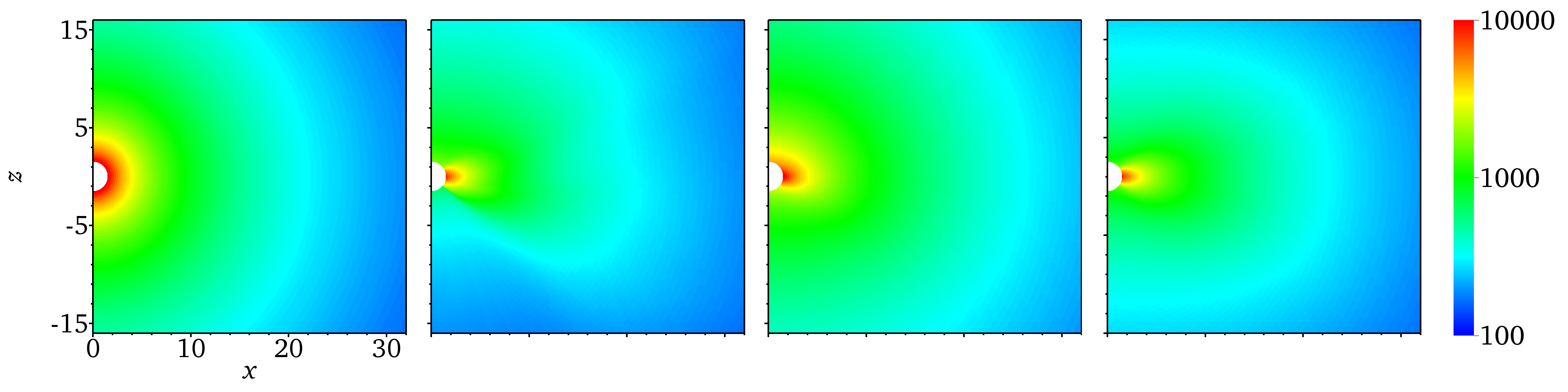}
\caption{\label{30_dens_fixed}Model 2b from Table \ref{table1}. The plots show, from left to right, a sequence in the evolution of the density $\rho$ at times $t = 0$, $25\times 10^5$, $50\times 10^5$, and  $75 \times 10^5$. The asymptotic value of the density is $\rho_\infty = 1$.}
\end{center}
\end{figure*}

\begin{figure*}
\begin{center}
\includegraphics[width=\textwidth]{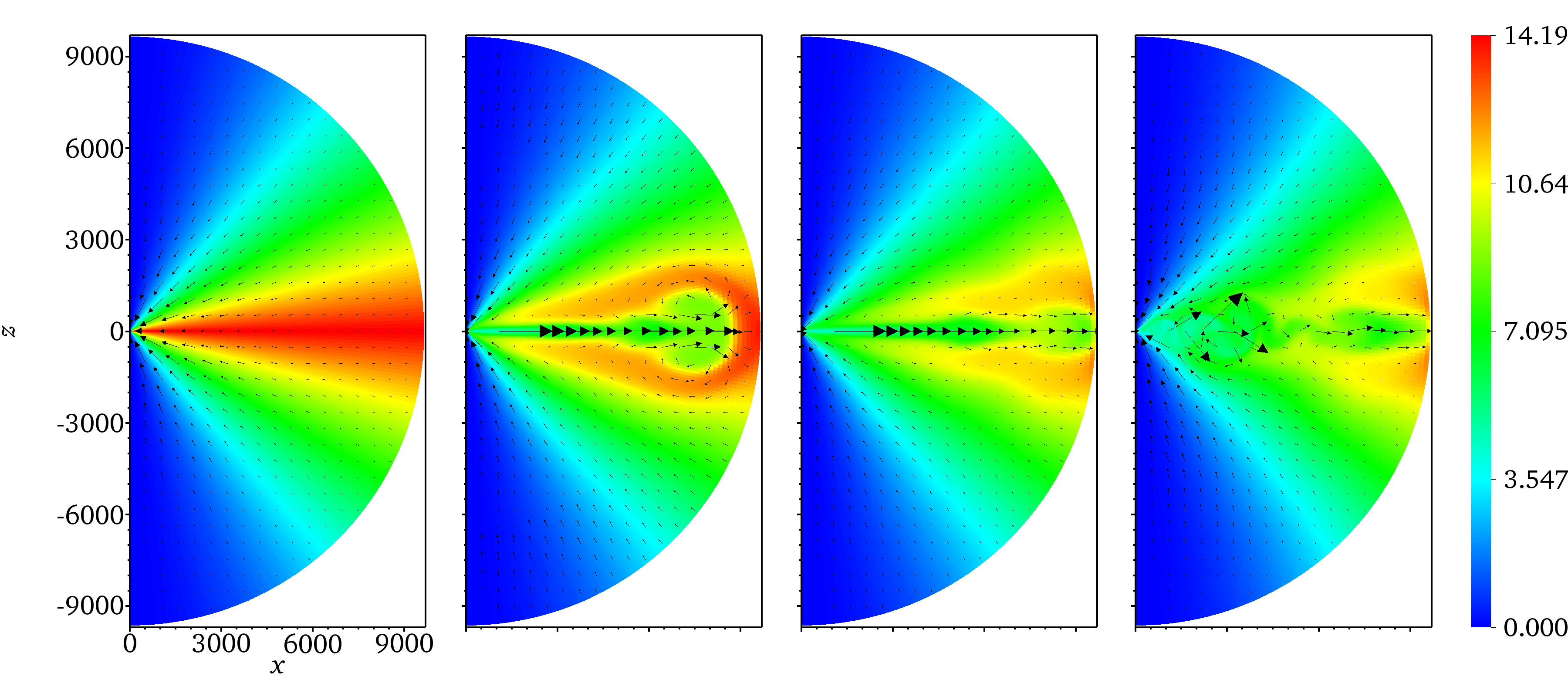}
\caption{\label{35_ang_free}Model 4a from Table \ref{table1}. The plots show, from left to right, a sequence in the evolution of the angular momentum (color coded) and of the velocity in the meridional plane (arrows) at times $t = 0$, $15 \times 10^5$, $30 \times 10^5$, and $40 \times 10^5$.}
\end{center}
\end{figure*}

\begin{figure*}
\begin{center}
\includegraphics[width=\textwidth]{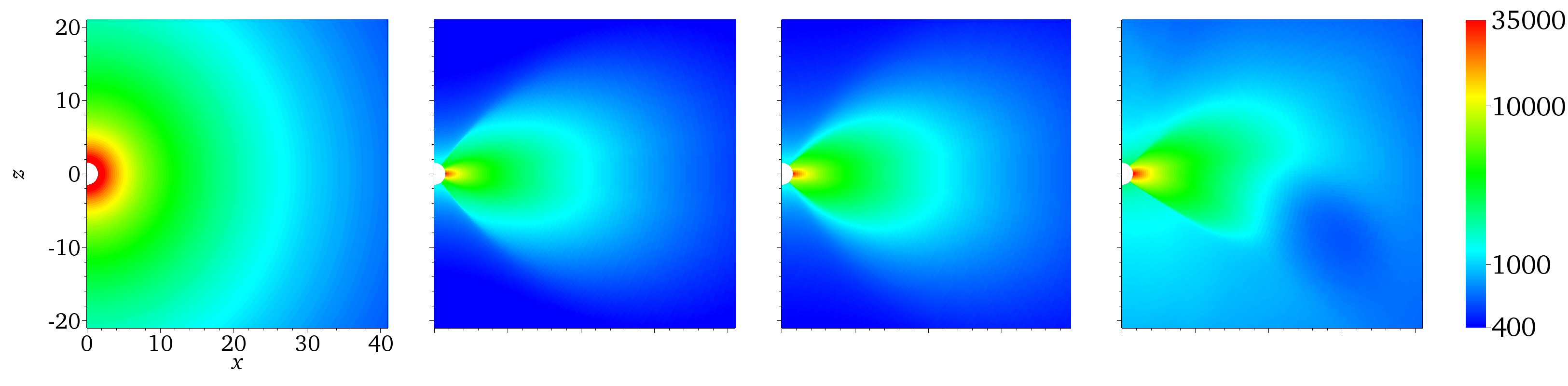}
\caption{\label{35_dens_free}Model 4a from Table \ref{table1}. The plots show, from left to right, a sequence in the evolution of the density $\rho$ at times $t = 0$, $15 \times 10^5$, $30 \times 10^5$, and $40 \times 10^5$. The asymptotic value of the density is $\rho_\infty = 1$.}
\end{center}
\end{figure*}

Figures \ref{20_ang}--\ref{35_dens_free} show examples of the evolution of (some of) the models reported in Table \ref{table1}. Figures \ref{20_ang} and \ref{20_dens} depict the evolution of model 1b from Table \ref{table1}. We do not consider this model as physically relevant as the rest of models due to relatively high value of the asymptotic speed of sound, $\sosi = 0.07071$. It is however important as a reference for models with lower values of $\sosi$. Figure \ref{20_ang} contains intensity plots of the specific angular momentum $l$ (color coded) at subsequent instants of time. The arrows in the plots correspond to the three-velocity field in the meridional plane. They are proportional to the vector field with the coordinates $(v_r - \beta_r)\sqrt{\gamma^{rr}}/\alpha$ and $v_\theta \sqrt{\gamma^{\theta \theta}}$. More precisely, we depict the vector field with the Cartesian coordinates (in the meridional plane) that are proportional to $(v_r - \beta_r)\sqrt{\gamma^{rr}}/\alpha \sin \theta + v_\theta \sqrt{\gamma^{\theta \theta}} \cos \theta$ ($x$ coordinate) and $(v_r - \beta_r)\sqrt{\gamma^{rr}}/\alpha \cos \theta - v_\theta \sqrt{\gamma^{\theta \theta}} \sin \theta$ ($z$ coordinate). Figure \ref{20_dens} shows the density in the vicinity of the central black hole. The accretion occurs in a broad region around the rotation axis, and it is stopped at the equator. This is connected with a slight distortion of the density profile, visible in the right panel of Fig.~\ref{20_dens}, which is no longer spherically symmetric. The model in Figs.\ \ref{20_ang} and \ref{20_dens} was computed for the outer boundary conditions with fixed $v^\theta = 0$. The corresponding model with ``free $v^\theta$'', not shown, evolves to a state that does not resemble an accretion flow.

The time evolution of the mass accretion rate computed for those two models is plotted in Fig.\ \ref{m20}. For model 1b (the one with the ``fixed $v^\theta$'' boundary condition) the mass accretion rate settles rapidly at about 0.7 of the rate that is characteristic for the Michel solution, and it exhibits quite regular oscillations. As we show below, there are no such oscillations for more realistic models with lower values of $\sosi$ (or they are hidden beneath larger ``turbulent'' changes). On the other hand, model 1a with a ``free $v^\theta$'' boundary condition shows a remarkable different behavior.  There is an initial transient phase in which the mass accretion rate first drops to then increase to reach a maximum value of $\sim 1.1\dot{m}_0$. This is followed by a steady  decline towards a long-term, small constant value of $\sim 0.2\dot{m}_0$.

Figures \ref{30_ang_free}--\ref{30_dens_fixed} depict the results obtained for models 2a and 2b from Table \ref{table1}. They are computed for $\sosi = 0.02236$ and $l_0 = 4.5714$ and show a behavior that we consider to be representative also for other models. The most characteristic feature of these models is a sudden symmetry breaking of the flow, followed by a ``turbulent'' evolution. This is best visible on the intensity plots of the specific angular momentum $l$ (Figs.\ \ref{30_ang_free} and \ref{30_ang_fixed}). Figure \ref{30_ang_fixed} shows the evolution of the model with fixed $v^\theta = 0$ at the outer boundary while Fig.\ \ref{30_ang_free} shows the corresponding evolution for the model with ``free $v^\theta$''. It is difficult to point out any qualitative difference between these two models, and this fact is also confirmed in Fig.\ \ref{m30}, where the time dependence of the mass accretion rate $\dot m$ is plotted. In both cases the accretion rate drops initially from the value that is characteristic for the unperturbed Michel solution ($\dot m_0$) to about $0.35 \dot m_0$, and then it grows slightly. The ``turbulent'' evolution of both models is reflected in highly irregular variations of the mass accretion rate in time.

In addition, both models 2a and 2b show the development of a geometrically thick, nearly regular disk around the equatorial plane. Sample plots of the density near the black hole are shown in Figs.\ \ref{30_dens_free} and \ref{30_dens_fixed} which clearly indicate the formation of such accretion torus. Note that the color-scale of the density is logarithmic in these plots.

Figures \ref{35_ang_free} and \ref{35_dens_free} show sample results obtained for the time evolution of model 4a from Table \ref{table1}. This is probably the most realistic case of our series, as compared to the models that have been discussed so far. In model 4a $\sosi = 0.01257$; this corresponds to the Bondi radius $R_\mathrm{B} = 6300 M$, and we set the outer boundary of the grid at almost $r_\mathrm{out} = 9500 M$. Similarly to models 2a and 2b, in the evolution of model 4a one also observes the breaking of the initial symmetry of the flow and a subsequent ``turbulent'' evolution at  late times (cf.~Fig.\ \ref{35_ang_free}). Figure \ref{35_dens_free} shows the formation and evolution of a geometrically thick accretion disk, with the shape that resembles the familiar structure of test-fluid stationary toroids computed on a fixed Schwarzschild or Kerr background.

The mass accretion rates for models 4a and 4b are plotted in Fig.\ \ref{m35}. This figure shows that the evolution of the mass flux is remarkably similar for the two models, and differences only become visible at very late times, after $t \sim 3 \times 10^6$. Both  of them show a pronounced drop of $\dot m$ initially, to about 0.2 of the Michel value $\dot m_0$, followed by a linear steady increase with time. This behaviour is identical to that found for models 2a, 2b, 3a, and 3b, depicted in Fig.\ \ref{m30}. By comparing the two figures we notice that the initial drop in $\dot m$ increases with increasing $l_0$ and with decreasing values of $\sosi$. The slope of the subsequent linear increase of $\dot m$ is smaller the larger $l_0$ and the smaller $\sosi$, and the final ``turbulent'' state, characterized by noticeable variations in the accretion rate with no definite pattern, takes longer to appear. In fact, for models 4a and 4b, ``turbulence'' is about to appear at the time when the simulation was stopped ($t\sim 4\times 10^6$; cf.~Fig.\ \ref{m35}). Likewise, large variations in $\dot m$ for the models 3a and 3b depicted in Fig.\ \ref{m30} would have most likely been initiated had these models been evolved beyond the final time they were evolved, $t\sim 8\times 10^5$. It is also interesting to notice the periodic drops found during the phase of linear increase of the mass accretion rate, visible in Figs.\ \ref{m30} and \ref{m35}, and whose time span increases with $l_0$. This feature is most probably due to reflections from the outer boundary of the numerical grid although, as it can be observed in Figs.\ \ref{m30} and \ref{m35}, it appears identically both for models ``a'' and ``b''. We have checked that the time span of the drops in the accretion rate increases considerably when the same calculation is repeated on a numerical grid with the outer boundary located at a much larger radius. In principle, both Figs.\ \ref{m30} and \ref{m35} can be compared with an analogous Fig.\ 5 from Ref.\ \cite{proga_begelman}, obtained in the Newtonian case.

\begin{figure}[t!]
\begin{center}
\includegraphics[width=0.75\textwidth]{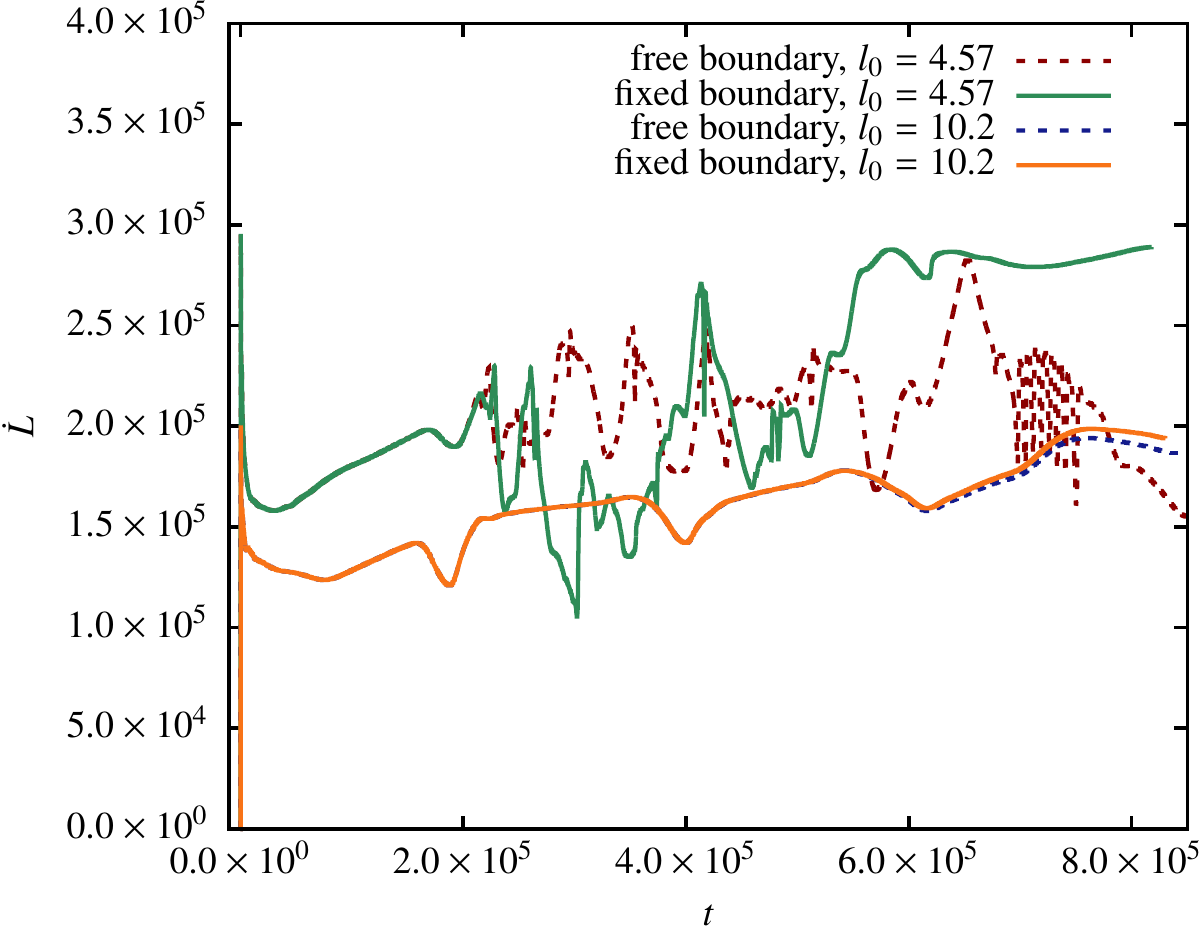}
\caption{\label{l30}Time evolution of the angular momentum accretion rate for models 2a (red dashed curve), 2b (green solid curve), 3a (black dashed curve), and 3b (orange solid curve).}
\end{center}
\end{figure}

\begin{figure}[t!]
\begin{center}
\includegraphics[width=0.75\textwidth]{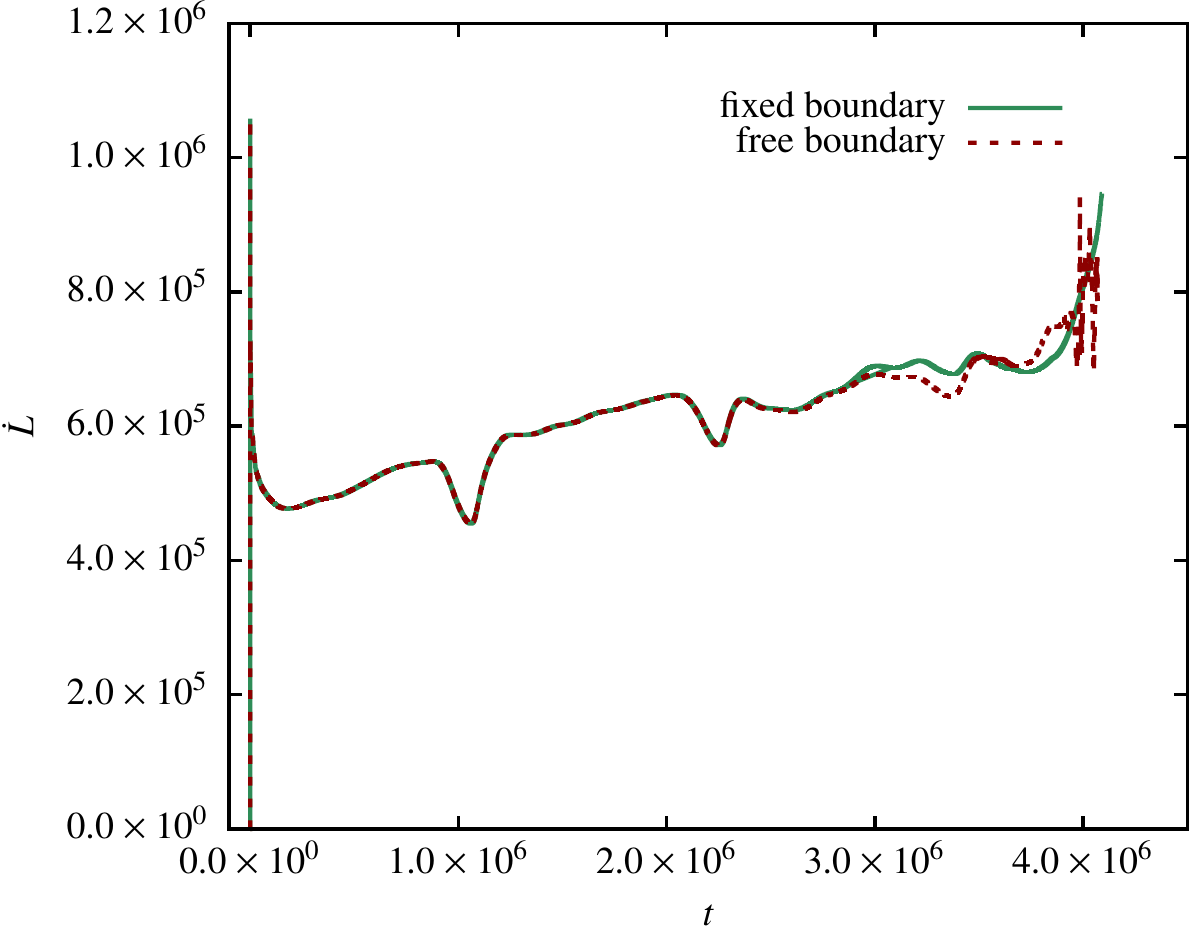}
\caption{\label{l35}Time evolution of the angular momentum accretion rate for models 4a (``free $v^\theta$'' at the boundary) and 4b (``fixed $v^\theta = 0$'' at the boundary).}
\end{center}
\end{figure}

Figures \ref{l20}--\ref{l35} depict the time evolution of the accretion rate of the angular momentum. The graphs in these figures are plotted for the same models for which the corresponding Figs.\ \ref{m20}--\ref{m35} show the mass accretion rate. Both rates are strongly correlated, and this is suggested already by Eqs.\ (\ref{ggh5}) or (\ref{ggh6}), where the integrand for the computation of $\dot L$ is proportional to $\dot m$. The oscillations in the mass accretion rate that are caused by the ``turbulent'' behavior of the accreting gas are also reflected in the time dependence of the accretion rate of the angular momentum.  Of course, in our setup, the starting value (for $t = 0$) of $\dot L$ is zero, whereas for $\dot m$ it is precisely $\dot m_0$.

It is also worth to compare the accretion rates of the angular momentum for models with different initial and boundary angular momentum (the parameter $l_0$). On the one hand one would expect that with increasing angular momentum, there is more angular momentum available for accretion. On the other, increasing angular momentum leads to a larger centrifugal barrier. A comparison of the models that have been computed shows that the accretion rate of the angular momentum decreases slightly with increasing $l_0$ (e.g.\ for models with $\sosi = 0.02236$, $l_0 = 4.57$, 10.2).

The accretion rates discussed here were computed on the horizon ($r = 2M$). It is also interesting to check the behavior of mass accretion rates computed at the outer boundary of the numerical grid. We expect that they should reflect inadequacy of the outer boundary conditions with respect to the physical situation (properties of a quasi-stationary flow). In our case the observed variation of the mass accretion rate at the outer boundary of the grid is much larger than that at the horizon. We obtain ratios as high as $|\dot m/\dot m_0| \sim 3$ for model 2a. This means that our simulations do not give reliable predictions of the physical flow in the outer regions of the numerical grid, where the influence of the boundary conditions is strong. On the other hand the simulated accretion flow in the vicinity of the black hole seems to be weakly affected by the unaptness of the outer boundary conditions.

In this paper we generally adopted the normalization with the asymptotic value of the density $\rho_\infty = 1$ (in consistency with the previous Newtonian studies). That of course means that the mass of the fluid contained within the boundaries of the numerical grid is much larger than the mass of the central black hole. Because we consider non self-gravitating models, the main properties of the Michel flows are only affected by the asymptotic speed of sound $c_{\mathrm{s},\infty}$. The situation is different for self-gravitating gases. Models of relativistic Michel-type accretion which take into account the self-gravity of the fluid were investigated in \cite{m1999, kkmms, mm, dok}.

\section{Conclusions}
\label{sec_conclusions}

In this paper we have carried out a numerical study of relativistic low angular momentum accretion of inviscid perfect fluid onto a Schwarzschild black hole. We have constructed our initial model setup as similar as possible to that of~\cite{proga_begelman}, in order to compare with those earlier Newtonian simulations. This is the reason why we have focused in this work on presenting the results of our axisymmetric simulations. The main difference between our setup and that of~\cite{proga_begelman} has been the use of relativistic gravity and relativistic hydrodynamics in our setting. In addition, we have taken advantage of the coordinate freedom provided by general relativity and have used horizon-penetrating coordinates. This allows us to locate the inner boundary of our grid within the black hole horizon, which removes potential sources of inaccuracies in regions of strong gravity near the black hole horizon. We have adapted the code of~\cite{font_bondi_hoyle} to perform our numerical simulations of low angular momentum accretion flows and have shown the theoretical second-order accuracy of the resulting code using the spherically-symmetric Michel solution and axisymmetric, stationary Fishbone-Moncrief toroids as tests.

It is worth to emphasize the conceptual simplicity of the models we have studied. They are purely hydrodynamical and basically consist of a background spherically symmetric Bondi (or Michel) flow to which we add a small amount of angular momentum. Such minimum setup turns out to be sufficient to establish a quasi-stationary, ``turbulent'' state with a relatively low mass accretion rate. We should note that taking into account the effects connected with viscosity could impact the observed ``turbulent'' effects. General-relativistic viscous hydrodynamics is an active field of research, also in the context of accreting systems (cf.\ the appendix in \cite{shibata}).

Nevertheless, technical difficulties in performing reliable multidimensional simulations for such setup do exist and they are due to the spatial (and temporal) scales involved -- we need to resolve both the region in the vicinity of the black hole but also cover the entire accretion cloud to radii larger than the Bondi radius. This poses a considerable computational demand, especially for low asymptotic sound speed models with a large Bondi radius.

Our relativistic results show that the overall morphology and dynamics of low angular momentum accretion found by~\cite{proga_begelman} remain essentially unchanged, hence supporting those Newtonian findings. In both cases the mass accretion rate drops considerably below the level characteristic for the Bondi-type (Michel) flow. This is, of course, desired in the context of inactive accretion systems at the center of galaxies. For the kind of cold accretion clouds we have simulated (still much hotter than presumed real astrophysical systems, which remain out of reach of computational resources) the flow is ``turbulent'', and the numerical noise is sufficient to trigger such ``turbulence''. In all cases considered, but particularly for (more realistic) models with the lowest asymptotic sound speed, a geometrically thick disk is formed, due to the centrifugal barrier. Accretion still occurs in the polar regions, along a funnel in the vicinity of the rotation axis.

\section*{Acknowledgments}

This work has been partially supported by the Polish Ministry of Science and Higher Education Grant No. IP2012 000172, the Spanish Ministry of Economy and Competitiveness  (AYA2013-40979-P), (AYA2015-66899-C2-1-P) and by the Generalitat Valenciana (PROMETEOII-2014-069). Numerical computations were carried out with the supercomputer ``Deszno'' purchased thanks to the financial support of the European Regional Development Fund in the framework of the Polish Innovation Economy Operational Program (contract no. POIG. 02.01.00-12-023/08).

\section*{References}

\end{document}